# High PerformancePiezoelectric Devices Based on Aligned Arrays of Nanofibers of Poly[(vinylidenefluoride-co-trifluoroethylene]

Luana Persano[1,2,+], Canan Dagdeviren[3,+], Yewang Su[4,5,+], Yihui Zhang[4,5], Salvatore Girardo[1], Dario Pisignano[1,2,6], Yonggang Huang[5], John A. Rogers[3]

[1] National Nanotechnology Laboratory of Istituto Nanoscienze-CNR, Università del Salento, via Arnesano, I-73100 Lecce (Italy).

[2] Center for Biomolecular Nanotechnologies @UNILE, Istituto Italiano di Tecnologia, Via Barsanti, I-73010 Arnesano –LE (Italy).

[3] Department of Materials Science and Engineering, Frederick Seitz Materials Research Laboratory, and Beckman Institute for Advanced Science, University of Illinois at Urbana-Champaign, Urbana, IL 61801, USA.

[4] Center for Mechanics and Materials, Tsinghua University, Beijing 100084, China.

[5] Department of Civil and Environmental Engineering and Department of Mechanical Engineering, Northwestern University, Evanston, IL 60208, USA.

[6] Dipartimento di Matematica e Fisica "Ennio De Giorgi", Università del Salento, via Arnesano I-73100 Lecce (Italy).

[+] These authors contributed equally to this work.

Correspondence and requests for materials should be addressed to J.A.R. (email: rogers@uiuc.edu).

Multifunctional capability, flexible design, rugged lightweight construction, and self-powered operation are desired attributes for electronics that directly interface with the human body or with advanced robotic systems. For these and related applications, piezoelectric materials, in forms that offer the ability to bend and stretch, are attractive for pressure/force sensors and mechanical energy harvesters. In this paper we introduce a large area, flexible piezoelectric material that consists of sheets of electrospun fibers of the polymer poly[(vinylidenefluoride-*co*-trifluoroethylene]. The flow and mechanical conditions associated with the spinning process yield free-standing, three-dimensional architectures of aligned arrangements of such fibers, in which the polymer chains adopt strongly preferential orientations. The resulting material offers exceptional piezoelectric characteristics, to enable, as an example, ultra-high sensitivity for measuring pressure, even at exceptionally small values (0.1 Pa). Quantitative analysis provides detailed insights into the pressure sensing mechanisms, and engineering design rules. Applications range from self-powered micro-mechanical elements, to self-balancing robots and sensitive impact detectors.

An emerging development trajectory inelectronics focuses onportable and flexible devicesfor applicationssuch as those that involve integration with the human body,as health/wellness monitors, surgical tools, sensor networks, artificial muscles and engineered tissue constructs[1-4].In the field of robotics, similar technologies can be important inefforts to optimize human-like manipulation schemes for robust modes of interaction in complex daily environments, at home or at work[5].Flexible,rugged and lightweight constructionare keyrequirements, for both areas.Devices that exploitmechanical motions as natural sources of power can be particularly valuable [6-10].Likewise,precision tactile sensorsmight represent first steps toward realization of artificial, electronic skins that mimick the full, multi-modal characteristics and physical properties of natural dermal tissues, for potential uses in robotics and human healthcare alike. Advanced materials will becritical to progress, particularly for integrated arrays that offerhigh sensitivity in the low-pressure regime (<1kPa). Previous examples include flexible, matrix-type arrays of pressure or strain sensors based on conductive rubbers[11-13]. Although such sensors have simple designs, their performance is modest, with pressure sensitivities around 0.5 kPa[11] and response slopes, in terms of percentage changes in resistivity with pressure,of 0.05% $(kPa)^{-1}$ [12].Recentcomponent level studies indicate thatsensitivitiesas high as 0.4 Pa[14] and 3 Pa[15] can be achieved withtriboelectric sensors and with air-gap capacitors, respectively.These systemsare of interest for many uses, but they suffer from undesirable sensitivities to static electricity, stray capacitances and temperature changes.

Piezoelectric polymers are especially promising for devices with this type of functionality, becausethey can exploit deformations induced by small forces, through pressure, mechanical vibration, elongation/compression, bending or twisting[16].These materials combine structural flexibility, ease of processing, and good chemical resistance, with large sensitive areas, simplicity in device design and associated potential for low cost implementation. Current state-of-the-art pressure sensors based on piezoelectric polymers mainly rely on thin film geometries. In parallel plate configurations[17], or combined with transistors on polymide substrates, such films enable precise pressure sensing, at a level suitable for detecting the touch of a finger (~ 2

kPa)[18].Poly(vinylidenefluoride) (PVDF) and its copolymers have particularly attractive piezoelectric properties in these geometries. Theirplastic behaviour makes them suitable forhigh throughput processing[16,19]based on molding, casting, drawing and spinning. Their carbon-fluorine chemistryrenders them highly resistant to solvents, acids, and bases. A disadvantage is that achieving good performance (~kPa sensitivity) requires electrical poling to create maximum polarization in the direction orthogonal to the film plane[20,21].This process constrains engineering design options and, at the same time, requires multiple preparatory steps.

Emerging techniques in nanofabrication have the potential tooptimizepiezoelectric responses, expand the range of device structures that can be considered, andsimplify processing.For example, near and far fieldelectrospinning methods can produce piezoelectric nanofibers, in whichthe associated extensional forces and electric fieldsnaturally cause local poling and, by consequence, enhanced properties. Previous reports, however, describe fibers of this type only in the form of non-woven mats orisolated strands[22-26].In particular, previously reported arrays of fibers of PVDF are present as monolayers or few layer coatings, with modest degrees of alignment (± 20°) and separations of a few microns[24]. Furthermore, these fibers require high-electric fields for poling, provided by either post-processing [24] or near-field electrospinning [25]. Under comparable bending conditions, the current outputs of devices produced with such fibers are <5 nA and the voltages are in the range of 1-20 mV [24, 25]. Realistic device functionality undervarious stress conditions, such as compression and bending, can be best achievedwith high volumetric densities of aligned arrays of fibers. These high densities are difficult to obtain by means of gap collectors or near field spinning. Gap collectors generally have poor capabilities in producing large area samples[27] with high degrees of alignment and uniformity in coverage because of the critical dependence on residual charges on the fibers in the gap. On the other hand, near-field methods typically involve direct serial writing, which is unsuitable for realizing large area and multilayered aligned arrays of fibers.Here we describe procedures that allow this type of architecture, and demonstrate the key features using free-standing, high densityarraysas a piezoelectric textile that can coverlarge areas (tens of cm$^2$). The resulting materials are mechanically robust and can be handled easily, with the capability to be bent or twisted without fracture. Under bending conditions, these fibers exhibit currents up to 30 nA and voltage ≥ 1 V.Furthermore, detailed characterization and modelling studies indicate distinctive piezoelectric features, due to underlying alignment at the scale of both the polymer chains and the fibers. Flexible pressure sensors built simply by establishing electrical contacts to the ends of the aligned fibers show excellent sensitivity in the low pressure regime (0.1 Pa) and respond to both compressive and bending forces.Alternative device architectures allow other modes of use, in sensors of acceleration, vibration and orientation.

**Results**

**Free-standing arrays of aligned nanofibers.**

The optimal fiber spinning process occurs with a potential of 30kV between a nozzle tip with inner diameter of 200 μm, fed by a syringe pump at a flow-rate of 1mL/hr, and a collector at a distance of 6 cm. See Fig. 1. A unique feature of this setup is a collector disk with sub-cm width that rotates at angular speeds as high as 4000 rpm, corresponding to linear speeds > 16 m/s at the collector surface.The consequence of this arrangement is that it forces overlapping fibers, which yields mesoscopic joints and significantly enhanced mechanical robustness. Using a high boiling point (i.e. slowly evaporating) solvent such as DMF (Tb: 154°C) helps to ensure the formation of such joints as shown in Fig. 2. Joint lengths range from hundreds of nanometers to tens of micrometers. In most of cases, fibers are arranged in groups where mutual adhesion through formation of these joints takes place (Fig. 2a and 2b). Occasionally, joined fibers remain strictly parallel, twist or cross at certain angles with respect to each other (Fig.2c,d,e). The present set-up enables fiber arrays with widths of 0.8 cm, lengths of 25 cm and thicknesses of 10-40 μm, depending on the spinning time duration (Fig. 1b-c).The averagefiberdiameter for materials studied hereis 260 nm, with a distribution that appears in Fig. 1d. Thearrayshave densities of~$2\times10^7$ fibers per mm$^2$ of cross-sectional area, andan overall porosity of 65%. The alignment is uniform over many centimeters (Fig.1e and Supplementary Fig. S1).Fast Fourier transforms (FFTs)[28] of scanning electron micrographs yield quantitative information on the degree of alignment, as shown in Fig. 1f. In the following, conventional, randomly oriented fiber mats, produced by spinning onto a static collector, and spin-cast thin films provide points of comparison.

X-ray photoelectron spectroscopy (XPS) indicates a ratio of co-polymerpoly[(vinylidenefluoride-*co*-trifluoroethylene][P(VDF-TrFe)] of 0.73:0.27,in agreement with the ratio before electrospinning (Supplementary Fig. S2 and Supplementary Note 1).P(VDF-TrFe) generally exhibitsgoodpiezo- and ferro-electric behaviour and a single all-trans polar crystalline phase (β-phase)that is stable at room temperature[29].The ferroelectricity stems from electrical dipoles created by hydrogen and fluorine atoms in the VDF molecules, which are positioned perpendicularly to the polymer backbone[30].X-ray diffraction (XRD) patterns provide information on the long-range order and the crystal structure of both aligned arrays and random networks (Fig.3a, b).The results indicate that the overall crystallinity of the material in the aligned arrays is ~ 48% , while that for the random mats is 40%. The former value is, in fact, comparable to that of the best previous results producedin thin films by stretching[31]. Adistinguishing feature of the fiber arrays is that the rotating collector significantly enhances the fraction of the polar β–phase, as is shown in the X-ray results of Fig. 3a, b and in the polarized Fourier transform infrared (FTIR) spectraof Fig.3c and d. The polarβ–phase bands[32-34]appear distinctlyat 508, 846, 1285, and 1431 $cm^{-1}$. By contrast, the bands of the non-polar α–phase (532, 612, 765, 796, 854, 870, 970 $cm^{-1}$) are not appreciable.The fraction of β–phase determined by analysis of these FTIR results[31] is close to 85%. Furthermore,the spectrashow a significant increase inthe intensity of bands associated with vibrations that depend on chain orientation(1076 and 1400 $cm^{-1}$)[35]for light polarizedalong the longitudinal axis of the fiber (blue line in Fig.3d),and a corresponding reduction in the intensity of bands sensitive to dipolar orientation (508, 846, 1285, and 1431 $cm^{-1}$).These results suggest a preferential alignment of the main molecular chains along the fiber longitudinal axisand, at the same time, an enhancement of the orientation of piezoelectric active dipoles (C-F) in the direction perpendicular to this axis(Supplementary Fig. S3a). This finding is consistent with a preferential dipolar alignment along the direction of the electrospinning field during fiber collection (Fig. 1a).FTIR spectra collected under different polarization conditions highlight an exceptional level of alignment, even compared with previous experiments using rotating collectors,[26] for which the fibers exist in moderately aligned mats. For instance, the ratio of intensities of absorption bands at 846 cm$^{-1}$ and 1285 cm$^{-1}$[35] which are indicative of the symmetric stretching vibration of the piezoactive (C-F) dipoles, under different polarizations are 2 and 2.6 respectively, about two times higher than correspondingratios inprevious reports[26].This result is attributable to the combined effects offour-fold higher rotational speed (namely, 2.3-times higher linear speed)of our collector, which causes tighter mutual alignment of nanofibers and concomitantly stronger stretching, and to excellent fiber alignment. We note that such polarization effects arenot observed in free-standing P(VDF-TrFe) films (SupplementaryFig. S3c),where preferential chain alignment is not expected, and in randommats, mainly due to the random alignment of fibers (SupplementaryFig. S3d)[We note that the mats showed higher degree of crystallinity compared to the films (40% vs 35%), consistent with expected effects of electrospinning.]

**Piezoelectric sensor: experimental and theoretical studies.**

These characteristics make fiberarrayspromising as building blocks for ultrasensitivepiezoelectric sensors. Devices, formed simply by establishing electrical contacts to the ends of a ribbon-shaped sample of fiber arrayson a flexible polyimide (PI) support (Kapton; thickness selected between 75 and 225 μm), reveal large response to even minute applied pressures (Fig. 4).To evaluate the sensitivityquantitatively, a soft elastomer(PDMS posts in Fig.4a)deliveredwell defined levels of pressure to the arrays, while the electrical response was measured.Data in Fig.4chighlight well-behaved, linear variations in the output voltage with pressure, for various values of the effective contact areasbetween 9 and 36 mm$^2$ (squares with sides$L_{eff}$; Fig.4b), with slopes between0.41 and 0.79 mV/Pa.At ranges of pressure between 0.4 and 2 kPa, the devices show further improved sensitivity, i.e. 1.1 V/kPa. For a given pressure (10 Pa, $L_{eff}$ = 3 mm), the output voltage does not change significantly with length of the fiber array, over a range between 2 and 8 cm).Even without sensitive voltmeters, this level of response enables accurate measurement of compressive pressures as small as 0.1 Pa(Fig.4d).These response slopes, which provide a measure of sensitivity that is independent of data acquisition systems, together with the observed broad dynamic range, provide unique operation compared to that provided by other sensors based on capacitive (0.55 kPa$^{-1}$ in units of relative capacitance, in a range of 0.5- 2 kPa)[15] and piezoelectric (0.1 mV/Pa, in a range of 0.01-30 Pa)[18] effects, as well as most force and tactile sensing methods that are currently available for in-hand manipulation in robotics[36].

The observed behaviors can be well explained by an analytic piezoelectric model. Consider a resultantprojected component of piezoactive dipoles along the longitudinal axis of the fibers,$x_3$, as induced by the deformation(Fig.4b). The fiber arrayare transverselyisotropic with elastic, piezoelectric, anddielectric constants $c_{ij}$, $e_{ij}$, and $k_{ij}$[37], respectively.For an applied compression -$p$ along the $x_1$ direction (normal to fibers, Fig.4a) over aneffective contact length $L_{eff}$(Fig.4a),the strain $\varepsilon_{11}$ and electric field $E_3$ along the poling direction are obtained from the constitutive relation $-p = c_{11}\varepsilon_{11} - e_{31}E_3$ and $D_3 = e_{31}\varepsilon_{11} + k_{33}E_3$ (see Supplementary Note 2 for details), where $D_3$ is the electric displacement along the poling direction. When the fiber arrays connected to a voltmeter, $E_3$ and $D_3$ are related to the measured voltage $V$ and current $I$ by $V=L_{eff}E_3$ and $I = -h_{PVDF}w_{PVDF}\dot{D}_3$, where $h_{PVDF-TrFe}$=20 μm and $w_{PVDF-TrFe}$= 8 mm are the thickness and width of the fiber array cross section, respectively. The voltage and current are then related by the resistance of the voltmeter, such that the voltage across $L_{eff}$ (see Supplementary Note 2 for details) is

$$V = \frac{\bar{d}}{\bar{k}} L_{eff}\, p, \qquad (1)$$

where $\bar{d} = e_{31}/c_{11}$ and $\bar{k} = k_{33} + e_{31}^2/c_{11}$. For $\bar{d}/\bar{k} = 0.14\ \text{V}\cdot\text{m/N}$, Eq. (1) agrees well with the experimental results shown in Fig.4b for a wide range of pressures$p$ and the three effective contact lengths $L_{eff}$ = 3, 4.5 and 6 mm used in the experiments. The value of $\overline{d}/\overline{k}$ reaches or exceeds those achieved in films with extreme stretching and poling (e.g., ~ 0.045-0.094 Vm/N[38]). Eq. (1) also suggests that the voltage is independent of the total length $L_{PVDF-TrFe}$ (Fig.4e) of the P(VDF-TrFe) fiber arrays, which is also consistent with experimental results of Fig.4c.

Additional behaviours were observed in dynamic bending experiments performed using fiber arrays on PI films with thicknesses between 75 and 225 μm. A flexural endurance tester (IPC, CK-700FET) subjected the samples to cycling bending tests at 1 Hz (Fig.5c,e) and 2 Hz (Fig.5d,f). During compression, the sample buckled to generate a bent shape (Fig.5a), with curvature consistent with simple mechanics considerations. The measurements showed a periodic alternation of positive and negative output peaks, corresponding to the application and release of the buckling stress, respectively (Fig.5b). The ranges of short-circuit current and voltage outputs were 6-40 nA and 0.5-1.5 V, respectively. Both responses increased with increasing PI thickness and with increasing bending frequency. The maximum current (40 nA) and voltage (1.5 V) were observed from fiber arrays on 225 μm thick PI substrates at 2 Hz. Tests of up to 1000 cycles of bending and relaxing revealed no significant changes in output voltage or current.

As with applied pressure, simple analytical models can account for the behaviours under bending (see Supplementary Note 3 for details). Under compression the PI substrate of length $L_{PI}$ buckles into a sinusoidal form represented by the out-of-plane displacement $w = A\left[1+\cos(2\pi x_3/L_{PI})\right]/2$, where the origin of coordinate $x_3$ is at the center of the substrate, and the amplitude $A$ is related to compression $\Delta L$ of the substrate by $A \approx (2/\pi)\sqrt{L_{PI} \cdot \Delta L}$ [39]. Here the critical compression to trigger buckling, ~1 μm for 150 μm-thick and 6 cm-long PI substrate, is negligible as compared to $\Delta L = 3$ cm in the experiments. The strain along the poling direction $x_3$ at the mid-plane of the fiber arrays is given by $\varepsilon_{33} = -w''(h_{PI}+h_{PVDF-TrFe})/2$, where $h_{PI}$ and $h_{PVDF-TrFe}$ are the thicknesses of PI substrate and P(VDF-TrFe) fiber arrays, respectively. Its strain $\varepsilon_{11}$ along $x_1$ and electric field $E_3$ along the poling direction are obtained from the constitutive relation $0 = c_{11}\varepsilon_{11} + c_{13}\varepsilon_{33} - e_{31}E_3$ and $D_3 = e_{31}\varepsilon_{11} + e_{33}\varepsilon_{33} + k_{33}E_3$, where $D_3$ is the electric displacement along the poling direction, and is a constant to be determined. For short-circuit current measurement between two ends of the P(VDF-TrFe) fiber arrays, the voltage across the length of P(VDF-TrFe) $L_{PVDF-TrFe}$ is zero, which, together with the above equations, gives

$$D_3 = \left[2\bar{e}\sqrt{\Delta L/L_{PI}}\left(h_{PI} + h_{PVDF-TrFe}\right)/L_{PVDF-TrFe}\right]\sin\left(\pi L_{PVDF-TrFe}/L_{PI}\right),$$

where $\bar{e} = e_{33} - (c_{13}/c_{11})e_{31}$ is the effective piezoelectric constant. The current $I$ is then obtained from $I = -h_{PVDF-TrFe}w_{PVDF-TrFe}\dot{D}_3$, where $w_{PVDF-TrFe}$ is the width of P(VDF-TrFe) fiber arrays (such that $h_{PVDF-TrFe}w_{PVDF-TrFe}$ is the cross section area). For a representative compression $\Delta L = \Delta L_{max}\left[1-\cos(2\pi t/t_0)\right]^2/4$ with the maximum compression $\Delta L_{max} = 3$ cm and period $t_0$=0.5 and 1 second as in experiments, the maximum current is given by

$$I_{max} = 2\pi(-\bar{e})\frac{(h_{PI} + h_{PVDF-TrFe})h_{PVDFTrFe}w_{PVDF-TrFe}}{L_{PVDF-TrFe}t_0}\sqrt{\frac{\Delta L_{max}}{L_{PI}}}\sin\left(\frac{\pi L_{PVDF-TrFe}}{L_{PI}}\right). \quad (2)$$

For $L_{PVDF-TrFe}$=2.5cm and three thicknesses of PI substrate $h_{PI}$=75, 150 and 225 μm as in experiments, Eq. (2) gives the range of $I_{max}$ 14~27 nA for $t_0$=0.5 s and 5.6~14 nA for $t_0$=1 s, while experiments give 10~33 nA and 6.5~26 nA for $t_0$=0.5 and 1 s, respectively. Here the effective piezoelectric constant for the fiber arrays is taken as $\bar{e} = -2.1$ C/m$^2$, which is larger than that for films (~-0.4 C/m$^2$)[37] because of the strong anisotropy of arrays due to their fibrous structure.

For voltages measurements, $V$ is no longer zero. The electric displacement becomes

$$D_3 = \left[2\bar{e}\sqrt{\Delta L/L_{PI}}\left(h_{PI} + h_{PVDF-TrFe}\right)/L_{PVDF}\right]\sin\left(\pi L_{PVDF-TrFe}/L_{PI}\right) + \left(\bar{k}/L_{PVDF-TrFe}\right)V.$$

The current $I = -h_{PVDF-TrFe}w_{PVDF-TrFe}\dot{D}_3$ is also related to the voltage $V$ and the resistance $R$ of the voltmeter by $I=V/R$, which gives $V/R = -h_{PVDF-TrFe}w_{PVDF-TrFe}\dot{D}_3$, or equivalently

$$\frac{dV}{dt} + \frac{L_{PVDF-TrFe}}{\bar{k}Rh_{PVDF-TrFe}w_{PVDF-TrFe}}V = 2(-\bar{e})\frac{h_{PI} + h_{PVDF-TrFe}}{\bar{k}}\sin\left(\frac{\pi L_{PVDF-TrFe}}{L_{PI}}\right)\frac{d}{dt}\sqrt{\frac{\Delta L}{L_{PI}}}. \quad (3)$$

For $\Delta L = \Delta L_{max}\left[1-\cos(2\pi t/t_0)\right]^2/4$ and the initial condition $V(t=0)=0$, the maximum voltage is given by

$$V_{max} \approx 2\pi(-\bar{e})\frac{R}{t_0}\frac{(h_{PI}+h_{PVDF-TrFe})h_{PVDF-TrFe}w_{PVDF-TrFe}}{L_{PVDF-TrFe}}\sin\left(\frac{\pi L_{PVDF-TrFe}}{L_{PI}}\right)\sqrt{\frac{\Delta L_{max}}{L_{PI}}}. \qquad (4)$$

Forthree thicknesses of PI substrate $h_{PI}$=75, 150 and 225 μm as in experiments anda resistance of the voltmeter $R$= 70 MΩ, consistent with values resulting from independent resistance measurements,eq. (4) gives a range of $V_{max}$0.81~2.1 V for $T$=0.5 s and 0.29~0.85V for $t_0$=1 s, while experiments give 0.79~1.3V and 0.49~1.0V for $t_0$=0.5 and 1 s, respectively.

**Pyroelectric response.**

The pyroelectric response in these systems was also studied by measuring the current and voltage output upon heating/cooling cycles with a temperature range of 6 K around room temperature (SupplementaryFig.S4 and Supplementary Note 4).[40-42]The measured pyroelectric coefficient [41] for the fiber arrays was $\alpha = -68\,\mu C/(m^2 K)$ (SupplementaryFig. S5). For a constant heating/cooling rate 2.5 K/min (Supplementary Fig. S6), the maximum voltage measured from the voltmeter was 1.9 μV when the fiber arrays were in contact with the top of the heater (SupplementaryFig. S4b). The maximum voltage obtained from an analytic pyroelectric model is

$$V = -\alpha R h_{PVDF-TrFe} w_{PVDF-TrFe}\frac{dT}{dt}, \qquad (5)$$

which was also given by Lubormisky et al.[41], and is also shown in Supplementary Note 5. Equation (5) gives 1.81 μV for $dT/dt$=2.5 K/min and $\alpha$, $R$, $h_{PVDF-TrFe}$ and $w_{PVDF-TrFe}$ in the experiments (see in Supplementary Note 5 for details). This value is in good agreement with 1.9 μV measured in experiments. For the same set of parameters, Eq. (5) indicates that, at heating rates of ~6.5K/min,the pyroelectric voltage is only 5.6% of the piezoelectric voltage (0.084 mV) at the lower end of the range of pressure sensitivity (0.1 Pa, $\bar{d}/\bar{k} = 0.14$ V·m/N and $L_{eff}$=6 mm in Eq. (1) for the type of tests reported here).We note, however, that for most envisioned applications such as measurements of sound waves, pressure waves in arteries, mechanical vibrations, disturbances associated with breathing, motion in limbs, etc the timescales for mechanically induced change are much different than those associated with thermal processes.  Furthermore, in bio-integrated applications, the operating temperature range is narrow (<3 K), and the extent of mechanical deformation (in many cases) is large.  Such circumstances enable any necessary separation of mechanical and thermal signals by frequency filtering as part of backend data processing, with no change in the devices.

**Vibration sensor, accelerometer and orientation sensor.**

As demonstrators, we built devices capable of measuring vibration/acceleration and orientation. For the first, the fiber array serves as a diaphragm across a hole opened in a underlying plastic film, sealed over the closed cavity of a transparentbox (Fig.6a). Through the plastic film, movements are transmitted from the box frame to the array,which then operates asa vibrating mass of ~3 mg. Fig.6bdisplays the output voltage signal generated by this simple device, as response to environmental vibrations induced by sound pressure levels of 60-80 dB.The response included periodically alternating positive and negative voltage peaks, with a peak-to-peak output voltage that increasesfrom 6 to 14μV with sound intensity.These devices work in any orientation and can bemounted on any surface by means of transparent, skin-conformal plastic sheets (Fig.6c,d).  In a second example, a similar type of device, integrated on a solid support and with an attached test mass, acts as an orientation sensor.  Measurements on a inclined plane with variable angle, as sketched in the inset of Fig.6d, allow calibration of the response. Fig.6e showsthe output voltage collected at different inclinations, which yields a response that is consistent with a gravitation constant, $g$ = 9.8 m/s$^2$ [43], measured with experimental uncertainties of a few percent, mainly due to manual positioning of PDMS posts.

**Discussion**

Results presented here indicate that aligned P(VDF-TrFe) nanofibers can be formed into flexible, free-standing sheets, by use of electrospinning onto a fast rotating collector.  The process yields alignment at both at the level of the fibers and the polymers, thereby enabling excellent response and high piezoactive β-fractionwithout further processing (e.g. electrical poling). Combining experimental and theoretical approaches, both details of the material and device performances are presented under different operating conditions. Simple pressure sensors exhibit excellent response in the extremely small pressure regime. Other simple devices can be constructed easily, including accelerometers, vibrometers and orientational sensors. The collective results suggest utility in a variety of sensor and energy harvesting components, with lightweight construction, attractive mechanical properties and potential for implementation over large areas at low cost, with application opportunities in human motion monitoring and robotics.

**Methods**

**Electrospinning and electron microscopy.**

P(VDF-TrFe) (75/25 weight%, Solvay Solexis) was dissolved in 3:2 volume ratio of dimethylformamide/acetone (DMF/Acetone, Sigma Aldrich) at a polymer/solvent concentration of 21% w/w.Electrospinning was performed by placing 0.4-0.9 mL of solution into a 1.0 mL plastic syringe tipped with a 27-gauge stainless steel needle. The positive lead from a high voltage supply (XRM30P, 82 Gamma High Voltage Research) connected to the metal needle, for application of bias values around 25 kV. The solution was injected into the needle at a constant rate of 1 mL/h with a syringe pump (33 Dual Syringe Pump, Harvard Apparatus).A static collector made of a metallic plate covered with an Al foil, or a cylindrical collector (diameter = 8 cm, Linari Engineering S.r.l), was placed at a distance between 3 to 20 cm from the needle and biased at – 6 kV for the fabrication of random and aligned mats, respectively.To investigate the nature and structure of mesoscopic joints, arrays of fibers were made using solutions of PVDF-TrFe in Acetone, methylethylketone (MEK), Tetrahydrofuran (THF) and DMF/Acetone at volume ratios from 4:1 to 1:4. We found that joints form only when using DMF/Acetone. Continuous, free-standing, high dense arrays of fibers can be achieved for DMF/Acetone volume ratios in the range from 1:2 to 3:2. At the lowest DMF content, needle clogging interrupts electrospinning and thus limits the density of the arrays to $1 \times 10^3$ fibers per mm. At highest DMF content, fibers are discontinuous due to the presence of beads and necks. Here, the degree of mutual alignment is also strongly reduced. Extensive electrospinning experiments were also performed with Acetone, THF and MEK by varying the PVDF-TrFe /solvent concentration in the range 12-21% (w/w). The results, which appear in Supplementary Fig. S7, suggest that use of DMF is critically important in the formation of continuous, smooth fibers at high densities and with mesoscopic joints at crossing points. For purposes of comparison, PVDF fibers formed with the same experimental conditions were examined, to compare the morphological, crystallographic and mechanical properties against those of PVDF-TrFe (Supplementary Fig.S8 and Fig.S9).Films of P(VDF-TrFe) with thicknesses of 10-40 μm were deposited by spin-coating at 800 rpm. All the fabrication stepswere performed at room temperature with air humidity of about 40%.The morphological analysis was performed by scanning electron microscope (SEM) with a Nova NanoSEM 450 system (FEI), using an acceleration voltage around 5 kV and an aperture size of 30 μm.

For quantitative analysis of alignment, SEM micrographs were converted to 8-bit grayscale TIF files and then cropped to 880×880 pixels. ImageJ software (NIH, http://rsb.info.nih.gov/ij) supported by an oval profile plug-in (authored by William O'Connell) was used for radial summation of pixel intensities. All FFT data were normalized to a baseline value, and FFT images were rotated by 90° for better visualization.

**X-ray and spectroscopic characterization.**

XPS spectra of fibers were collected using a Kratos Axis ULTRA X-ray photoelectron spectrometer with monochromatic Al Kα-excitation, 120 W (12kV, 10 mA). To reduce the effects of surface charging, the monochromatic source was operated at a bias voltage of 100 V. Data were collected using the low magnification (FOV1) lens setting with a 2 mm aperture (200 μm analysis area) and charge neutralizer settings of 2.1 A filament current, 2.1 V charge balance and 2 V filament bias. Survey spectra were collected at a pass energy of 160 eV and high resolution spectra were recorded using a pass energy of 40 eV. The data were fitted with Gaussian-Lorentzian line shapes.The binding energy scale was referenced to the aliphatic C 1s line at 285.0 eV. IR spectroscopy was performed with an FTIR spectrophotometer (Spectrum 116 100, Perkin-Elmer Inc.), equipped with an IR grid polarizer (Specac Limited, U.K.), consisting of 0.12 μm wide strips of aluminum. The 4 mm wide beam, incident orthogonally to the plane of the sample, was polarized alternatively parallel or orthogonal to the main axis of fiber alignment.FTIR measurements performed on different samples and on different points of the same sample yielded similar results.A PANalyticalX'pert

MRD system, with Cu k-alpha radiation (wavelength 0.15418 nm), crossed-slit collimator as primary optics, and secondary optics consisting of a parallel plate collimator, a flat graphite monochromator and a proportional detector, was used for XRD measurements. A detailed scheme of the XRD set-up we used is reported in Supplementary Fig. S10. The crystallinity of the fibers was calculated from the area of the diffraction peaks (above the background) divided by the area of the whole diffraction curve. During X-ray analysis, samples were mounted on a low-background quartz holder. The instrument contribution to the background in the diffraction data was determined by separate measurements of the quartz holder without samples.

**Pressure sensor fabrication and characterization.**

P(VDF-TrFe) aligned fiber arrays were placed on 75-150 and 225 μm thick kapton film, and electric connections were established withcopper films (25 μm thick)and silver paint (Ted Pella Fast Dring Silver Paint, 160040-30). Open loop voltage measurements were performed by using a DAQ (SMU2055) USB multimeter (6.5 digit resolution, Agilent Technologies) with input resistance of R= 70 MΩ. Short-circuits current measurements were performed with a Semiconductor Parameter Analyzer (4155C Agilent Technologies) that has 10 fA measurement resolution. Signals were not amplified before acquisition.Pressure tests were performed usingPDMS posts with calibrated weights in the milligramrange, formed by replica molding against photolithographically defined templates. A Vacuum Pick-up Pen with Bent Metal Probe, 1" Long, 3/32" Cup diameter (Ted Pella, Inc., Vacuum Pickup System, 115V) was used to place the PDMS on the fiber arrays from calibrated distance to apply desired pressures. Upon impact, the PDMS postsadheredconformably and instantaneously, resulting in registered voltage spikes from the fiber array, digitized at 50 samples per second by the measurement system.These procedures allowed application of controlled pressures, in the low pressure regime (0.1- 12 Pa) onto an active P(VDF-TrFe) fiber array area of 9-18 mm$^2$.(Conventional load cells are difficult to implement for measurements in this range of pressures and areas.) We performed control experiments indicating that no significant signal is observed without the fibers and that interchanging the connections reversed the polarity of the output (Supplementary Fig. S11).All measurements were performed at 20°C.

**Bending measurements.**

Cycling tests and bending experiments were performed by using IPC Flexural Endurance Tester (Model: CK-700FET). The two edges of the sample were fixed within two sliding stripes. The buckling radius measured from the middle of PI substrate (max curvature) is 74 mm.

**Accelerometer, orientation sensor and vibration sensor measurements.**

A Personal Daq/3000 Series 16-bit/1-MHz USB Data Acquisition System was used to collect voltage signals from P(VDF-TrFe) fiber-based accelerometer. Orientation measurements were performed by placing devices on horizontal surfaces configured at an range of inclined angles. These experiments uses a PDMS test mass of 43 mg with $L_{eff}$ = 6 mm. Upon impact, the PDMS mass adhered to the fibers, resulting in voltage spikes from the device. Experimental data reported in Fig.4c were used to determine pressure values for voltage measurements at each angle.

**Acknowledgements**

Y. H. acknowledges support from ISEN, Northwestern University.D.P. and L.P. acknowledge the European Research Council for supporting, under the European Union's Seventh Framework Programme (FP7/2007-2013), the ERC Starting Grant "NANO-JETS" (grant agreement n. 306357). Mauro Sardela, Richard Haasc,Maria Moffa and Steve Burdinare acknowledged for support during XRD, XPS, DMA and accelerometer test respectively.L.P. thanks Eduardo Fabiano for molecular chain schematics.We thank J.S. Rogers for useful suggestions in accelerometer device design.

**Author Contributions**

L.P., J.A.R. and C.D. conceived experiments and device designs. L.P. and C.D.carried out experiments.Y.S., Y.Z. and Y.H. performed mechanical modeling. S.G. carried out preliminary experiments on solution and fibers processing. D.P. and J.A.R. provided further data analysis. L.P., C. D., Y.S., Y.H. and J.A.R. wrote the manuscript with input from all the other authors.All authors were involved in extensive discussions and data analysis.

**Additional information**

**Competing financial interests:** The authors declare no competing financial interests.

**Figure Legends**

**Figure 1 Arrays of highly aligned piezoelectric nanofibers of poly[(vinylidenefluoride-co-trifluoroethylene].(a)** Schematic illustration of the experimental setup for electrospinning highly aligned arrays of oriented nanofibers of aligned polymer chains of poly[(vinylidenefluoride-co-trifluoroethylene]. $\vec{E}$ indicates the direction of the electric field and $\Delta V$ is the applied bias. (**b**) Photograph of a free-standing film of highly aligned piezoelectric fibers. Scale bar: 1 cm. (**c**)SEM micrograph of fiber arrays intentionally folded many times to highlight flexibility and mechanical robustness (scale bar: 400 μm).(**d**) Typical fiber diameter distribution and fit by a Gaussian curve (solid line). (**e**) SEM micrograph of fiber arrays (scale bar: 10μm). (**f**) Radial intensity distribution vs detection angle (0-180°) for aligned arrays (wine line peak) and random mats (red line at bottom) of fibers. Insets: 2D FFTimages generated from aligned (left inset) and randomly oriented (right inset) fibers. In random mats, a highly symmetric, circular distribution of the pixel intensity confirms the un-oriented arrangement of fibers and correspondingly, a featureless behaviour of the radial pixel intensity vs angle. This distribution in fact, indicates that the frequency at which

specific pixel intensities occurs in the corresponding data image is identical in any direction and no peaks can be appreciated by plotting the sum of the pixel intensity as a function of the degree in the interval 0°- 360°.On the contrary, aligned fibers generate a highly anisotropic, elliptical 2D FFT profile with the major axis oriented parallel to alignment axis. Full width at half maximum of the radial intensity distribution = 16°.

**Figure 2 Mesoscopic inter-fiber joints**(**a**)SEM micrographs of fibers highlighting points of merging (joints) between two or more adjacent fibers and(**b**)their typical arrangement in groups where strong merging takes place (arrows indicate examples of joints in the micrographs. Scale bars: 3 μm and 2 μm, respectively). Magnification of crossed(**b**,scale bar: 1μm), adjacent(**c**,scale bar: 1μm) and twisted (**d**, scale bar: 500 nm) fiber geometries at the joints.

**Figure3Characterization of morphologies in properties of arrays.**(**a**) X-ray diffraction patterns (XRD; Cu radiation with wavelength = 0.15418 nm) from aligned fibers (top blue line), randomly oriented fibers (middle green line) and film (bottom red line) collected with the fibers length parallel to the diffraction plane (azimuthal rotation $\phi = 0$). The patterns are vertically shifted to facilitate comparison of the results. (**b**) Magnified view ofXRD spectra at $2\theta \cong 20°$ which corresponds to diffraction in (110) plane and represents the β-phase. Although the peak is visible in all samples (aligned arrays of fibers (blue dots), random mats of fibers (green dots) and bare films (red dots)), it is dominant in the aligned nanofibers arrays, where the contribution from α-phase is insignificant.(**c**) FTIR spectra measured under different incident beam polarizations for aligned fibers. Light is polarized parallel (orange line) and perpendicular (blue line) to the direction of the electric field used in electrospinning. Similar spectra were measured at different values of the electric field. (**d**) Magnified view of the low energy region of the FTIR spectra, highlighting the increase of the intensity of transitions sensitive to chain orientation (1400 cm$^{-1}$) for light polarized along the axis of the fibers (blue line).

**Figure 4 Experimental and theoretical studies of responses of pressure sensors.** (**a**) Photograph of the manipulator used to apply pressures, for the purpose of studying the voltage response. (**b**) Schematic illustration of an analytical model for the response of arrays of P(VDF-TrFe) fibers under applied compression -$p$ along $x_1$ direction over the effective contact length ($L_{eff}$). $L_{PVDF-TrFe}$ is the total length of the P(VDF-TrFe) fiber array. (**c**) Experimental (symbols) and theoretical (lines) pressure response curves at different $L_{eff}$. (**d**) Experimental (symbols) pressure response curve in the low pressure regime (0.1-1 Pa) at $L_{eff}$= 6mm. The line corresponds to a linear fit. (**e**) Experimental (symbols) and theoretical (lines) response at different $L_{PVDF-TrFe}$ (applied pressure = 10 Pa, $L_{eff}$= 3mm).

**Figure 5 Experimental and theoretical studies of responses of flexural sensors**. (**a**) Schematic illustration of an analytical model for the coupling of mechanical deformation and piezoelectric response during bending. $h_{PVDF}$ and $h_{PI}$ are the thicknesses of the P(VDF-TrFe) fiber array and of the PI substrate, respectively. $L_{PI}$ is the length of the PI substrate. Under compression the PI substrate buckles into a sinusoidal form represented by the out-of-plane displacement, $w$ and the amplitude $A$ which is related to compression $\Delta L$ of the substrate. (**b**) Measured voltage response of an array of P(VDF-TrFe) fibers under cycling bending at 1 Hz. The top and bottom insets show photographs of the device during bending and release, respectively. (**c, d**) Measured short-circuit output current and (**e, f**) voltage under dynamic bending tests at 1 Hz (left panels, **c, e**) and 2 Hz (right panels, **d**, **f**). Experiments used devices onto PI substrates with different thickness. From bottom to top, the PI thicknesses are 75 μm (red line), 150 μm (green line), 225 μm (blue line).

**Figure 6 Accelerometer and orientation sensor.** (**a**) Photograph of a simple, P(VDF-TrFe) nanofiber-based accelerometer. Scale bar = 1 cm. (**b**) Output voltage collected from this device

exposed to 70 dB sound intensity.(**c,d**) Photographs of a flexible device mounted on the skin of the arm and wrapped around a fiber.Scale bars = 2 cm. (**e**)Characterization of an orientation sensor, based on a pressure sensor with an attached test mass. The output voltage changes with orientation angle, $\theta$, in an expected manner.The inset provides a sketch of the device (yellow: PI; grey: fibers; blue: test mass) and the measurement geometry.

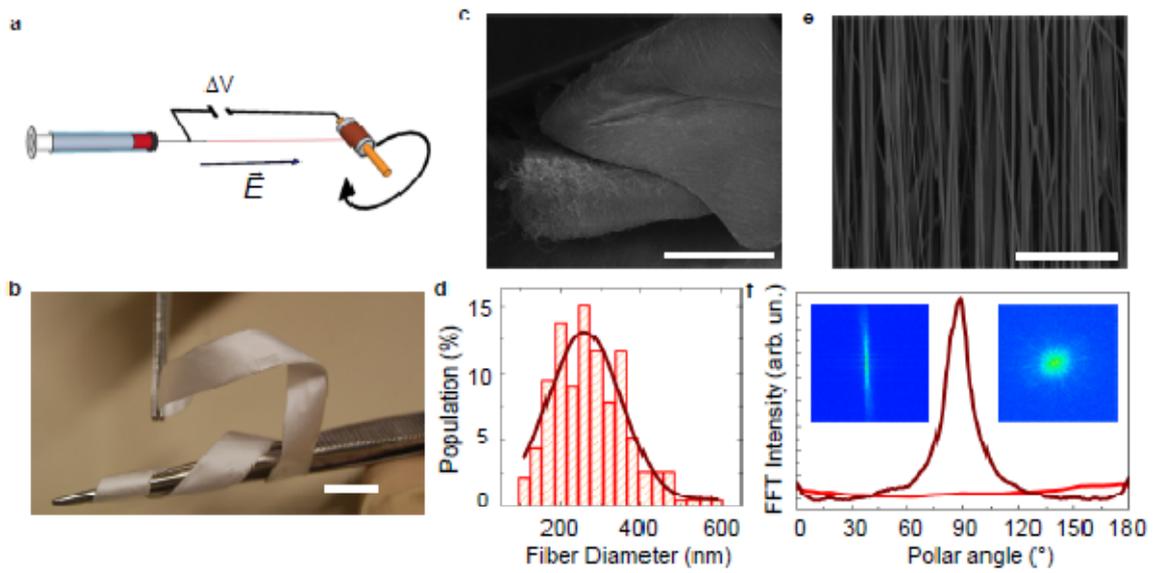

Figure 1

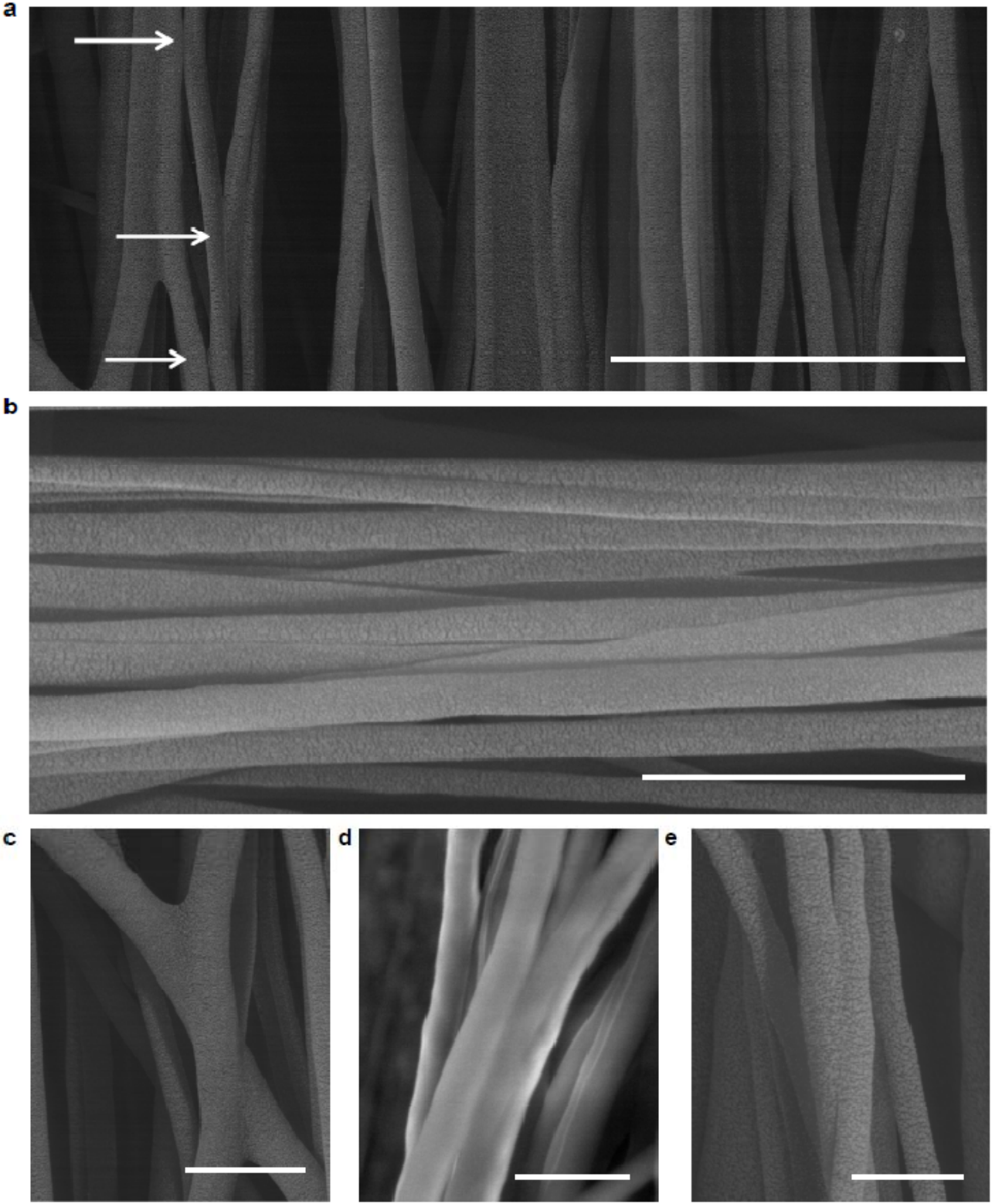

Figure 2

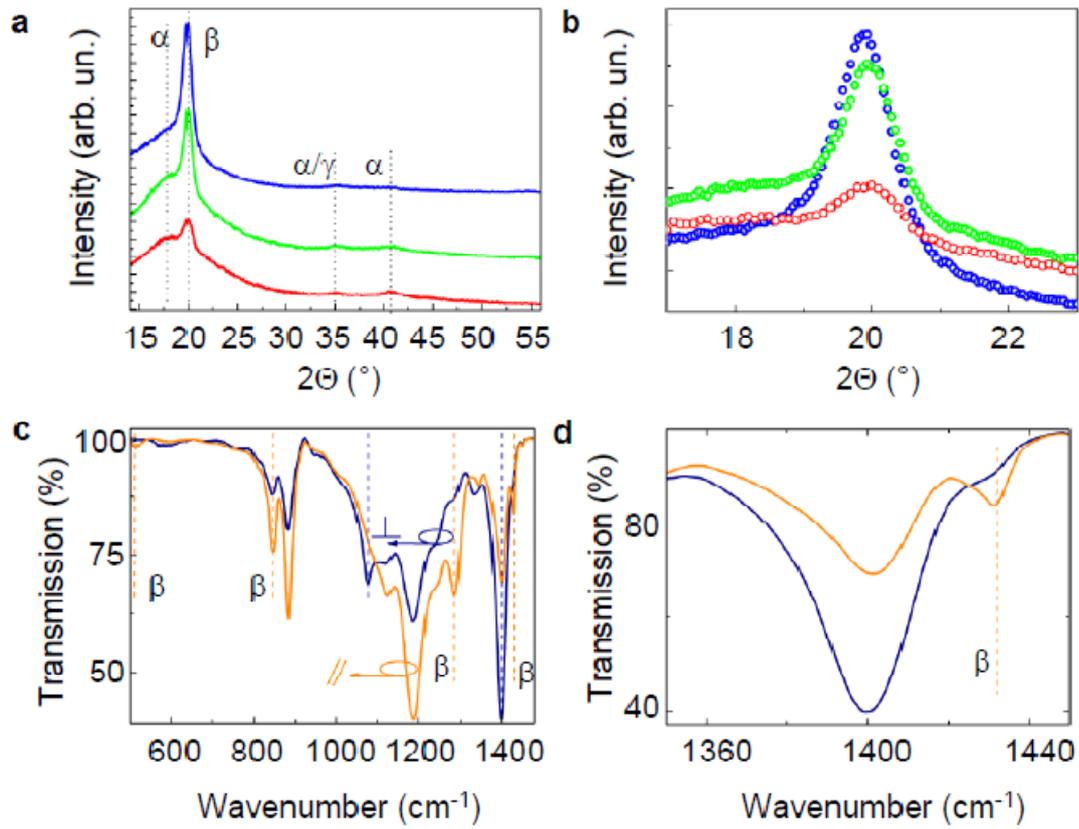

Figure 3

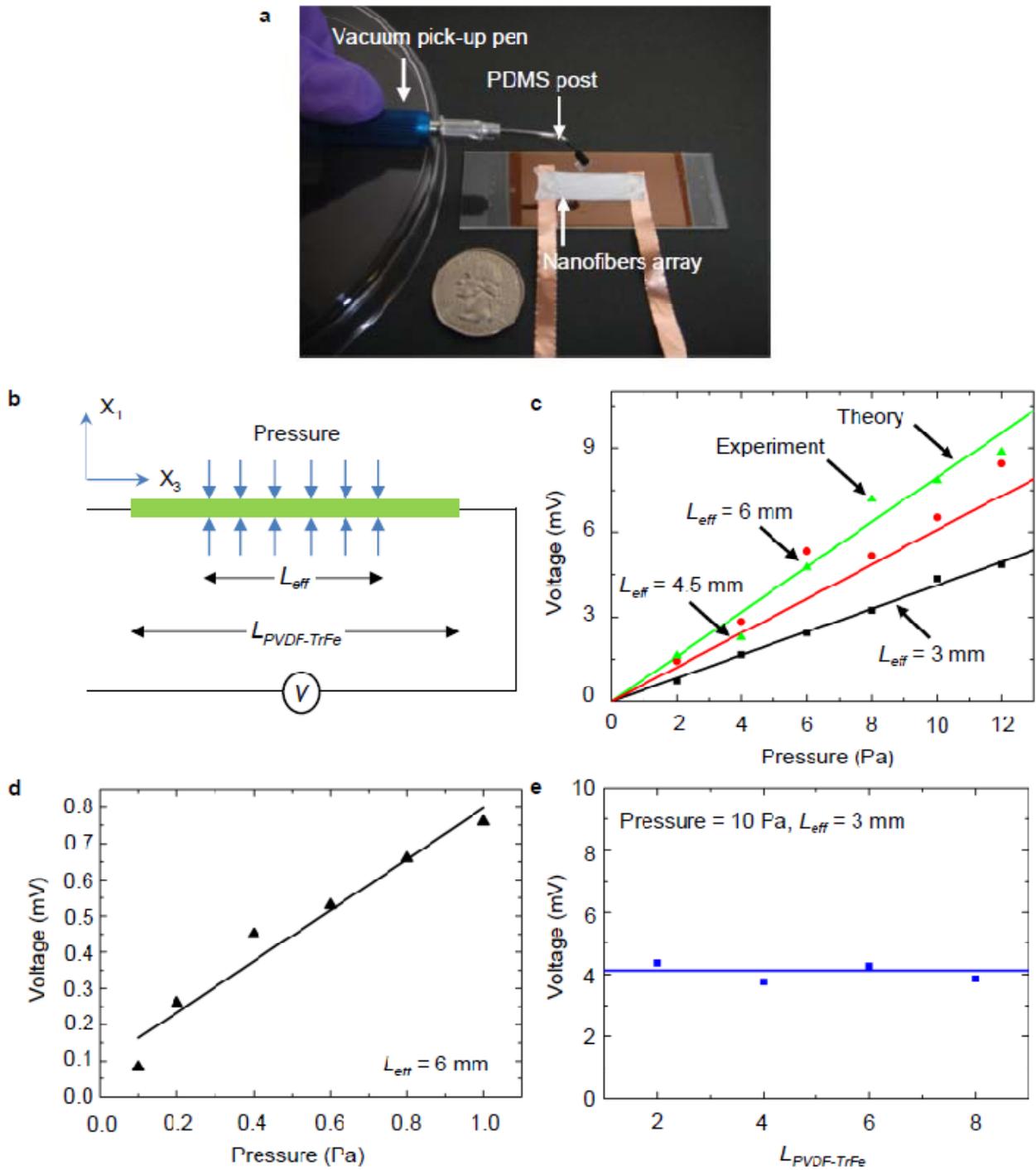

Figure 4

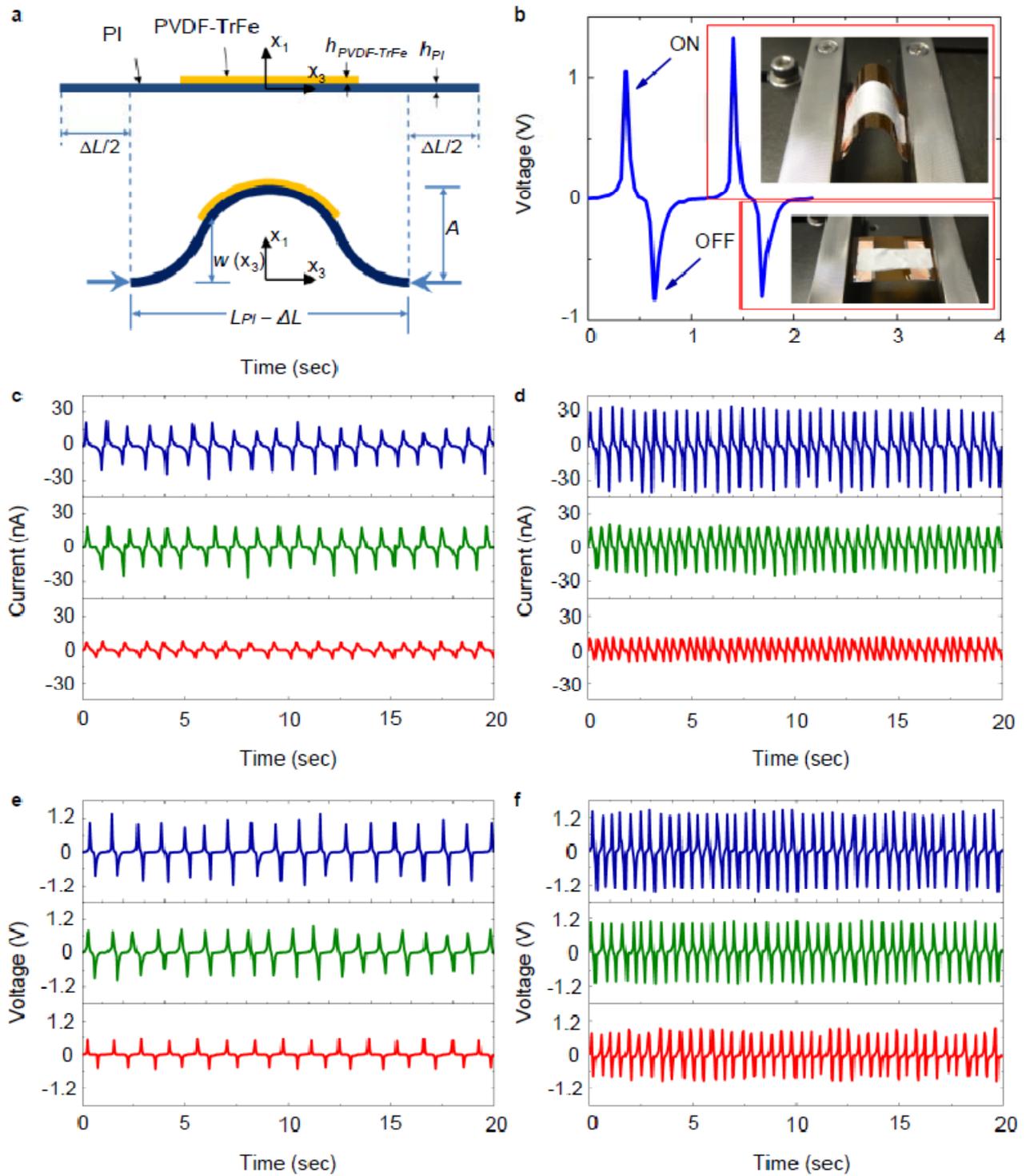

Figure 5

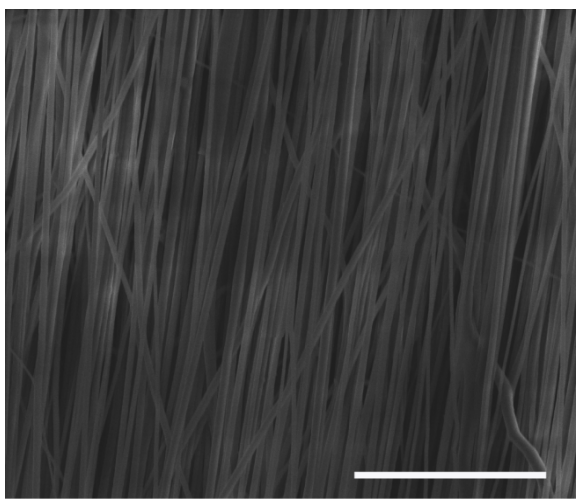

**Supplementary Figure S1│ High dense nanofibers array.** PrototypalSEM micrograph of high dense PVDF-TrFenanofibers array. Scale bar: 10 μm.

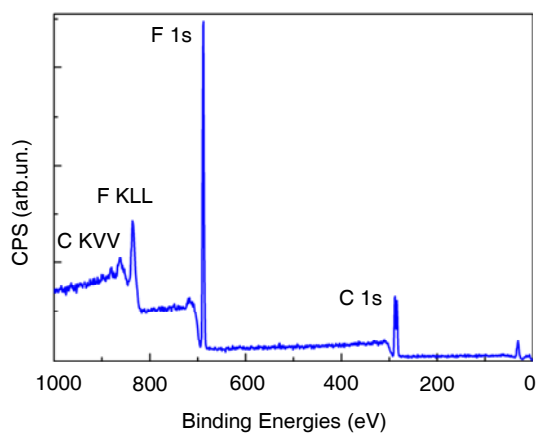

**Supplementary Figure S2│XPS on nanofibers array**. XPS plot of PVDF-TrFe nanofibers array collected using a Kratos Axis ULTRA X-ray photoelectron spectrometer with monochromatic Al Kα excitation, 120 W (12kV, 10 mA).

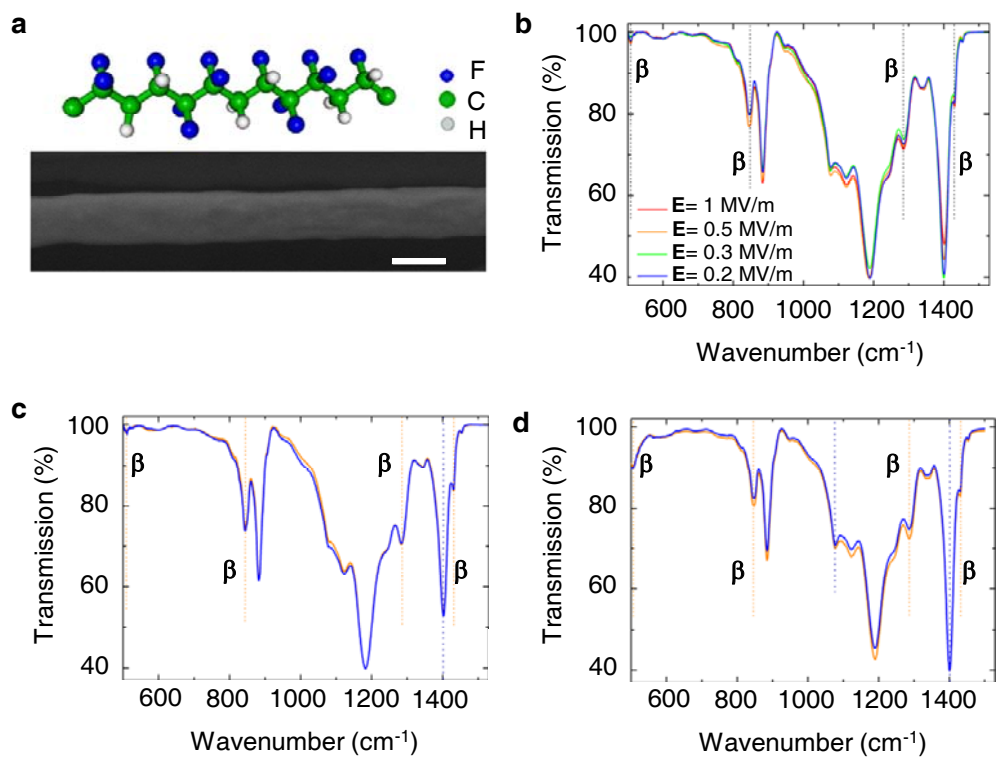

**Supplementary Figure S3│Nanofibers array characterization.** (**a**) Schematics of chain conformation of P(VDF-TrFe) crystal β-phase in nanofibers array, according to FTIR and XRD analysis. The main molecular chains are preferentially aligned along the fiber longitudinal axis and the piezoelectric active dipoles (C-F) are perpendicular to the backbone direction. Bottom: SEM micrograph of a single P(VDF-TrFe) nanofiber. Scale bar 250 nm (**b**)FTIR spectra of aligned fiber array, produced by different electric field during electrospinning (0.2-1 MV/m). **c-d,** FTIR spectra measured under two mutually orthogonal incident beam polarization direction for films (**c**), and random fibers produced by 0.5 MV/m (**d**).

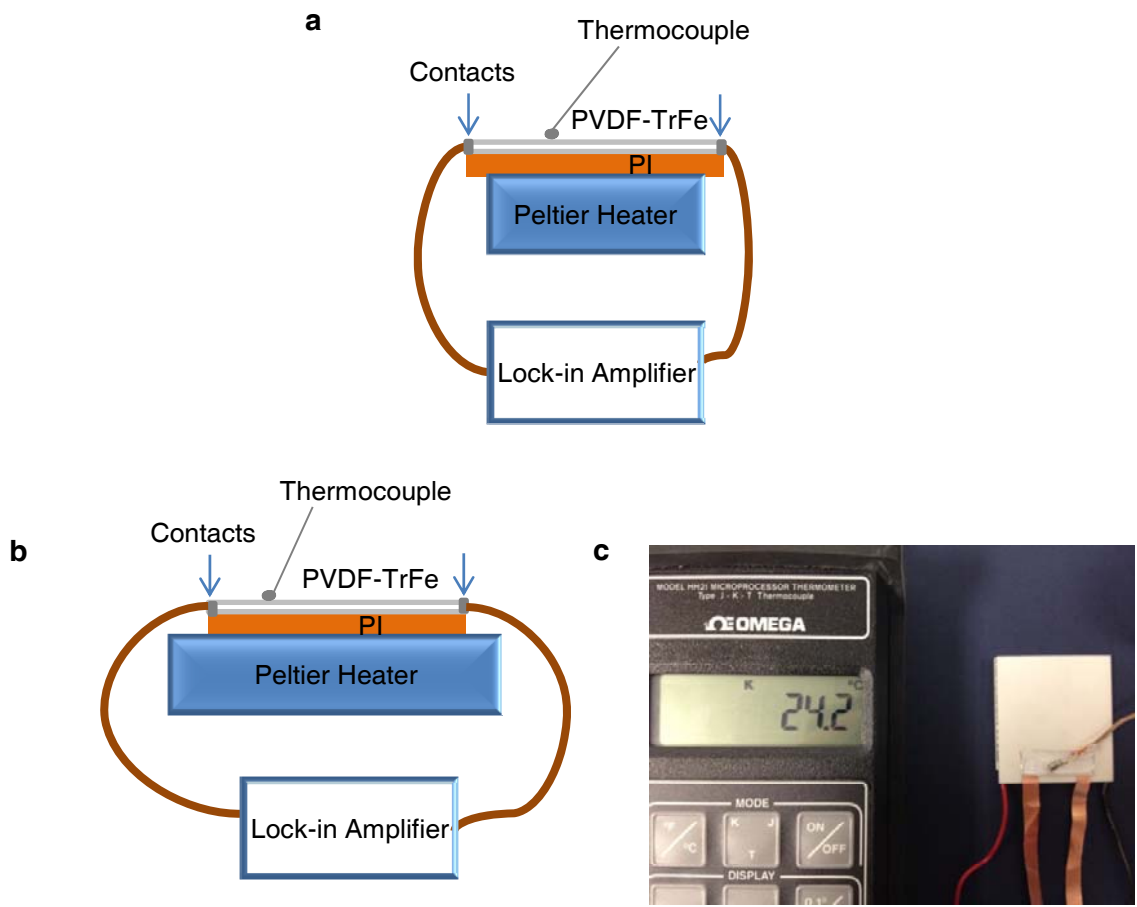

**Supplementary Figure S4│Set-up used to measure pyroelectric signals.** Schematic illustration of the setup used to measure the pyroelectric signal with contacts positioned outside of (**a**) and on top of (**b**) the Peltier stage. (**c**) Image of temperature evaluation during a typical pyroelectric measurement.

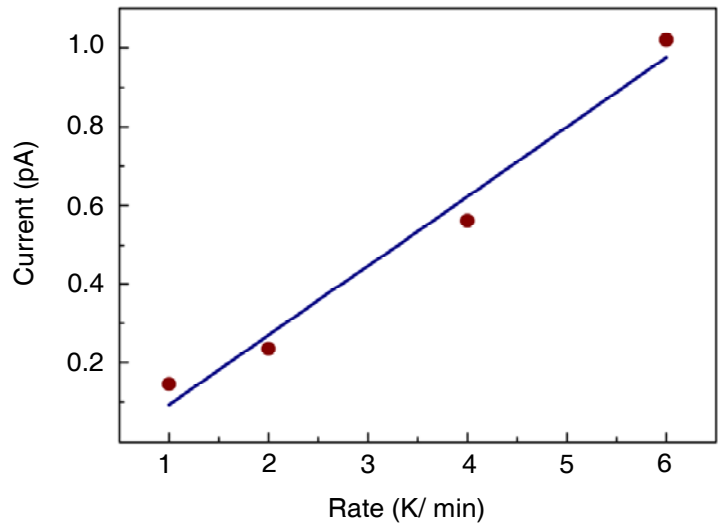

**Supplementary Figure S5│Pyroelectric current.** Plot of the pyroelectric current as a function of rate of temperature change (dots) at a temperature of 20°C. The line corresponds to a linear fit.

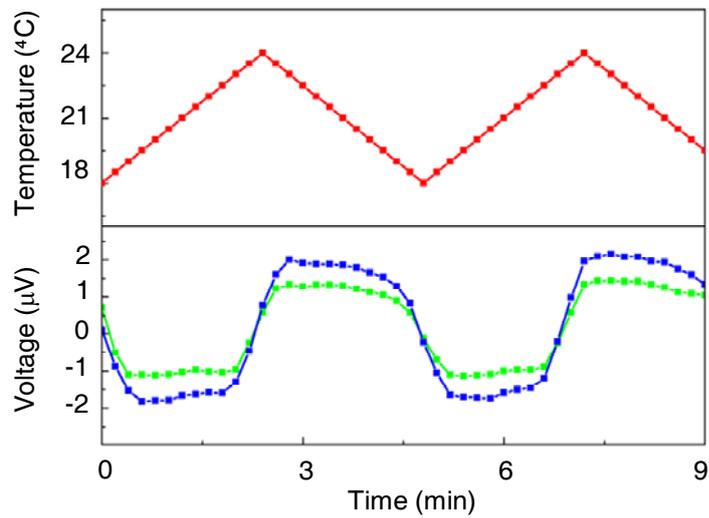

**SupplementaryFigure S6│Pyroelectric voltage.** Plot of the heating/cooling ramp applied to a nanofiber array device (red line) and the corresponding voltage recorded with contacts on top of (blue line) and outside of (green line) the Peltier stage.

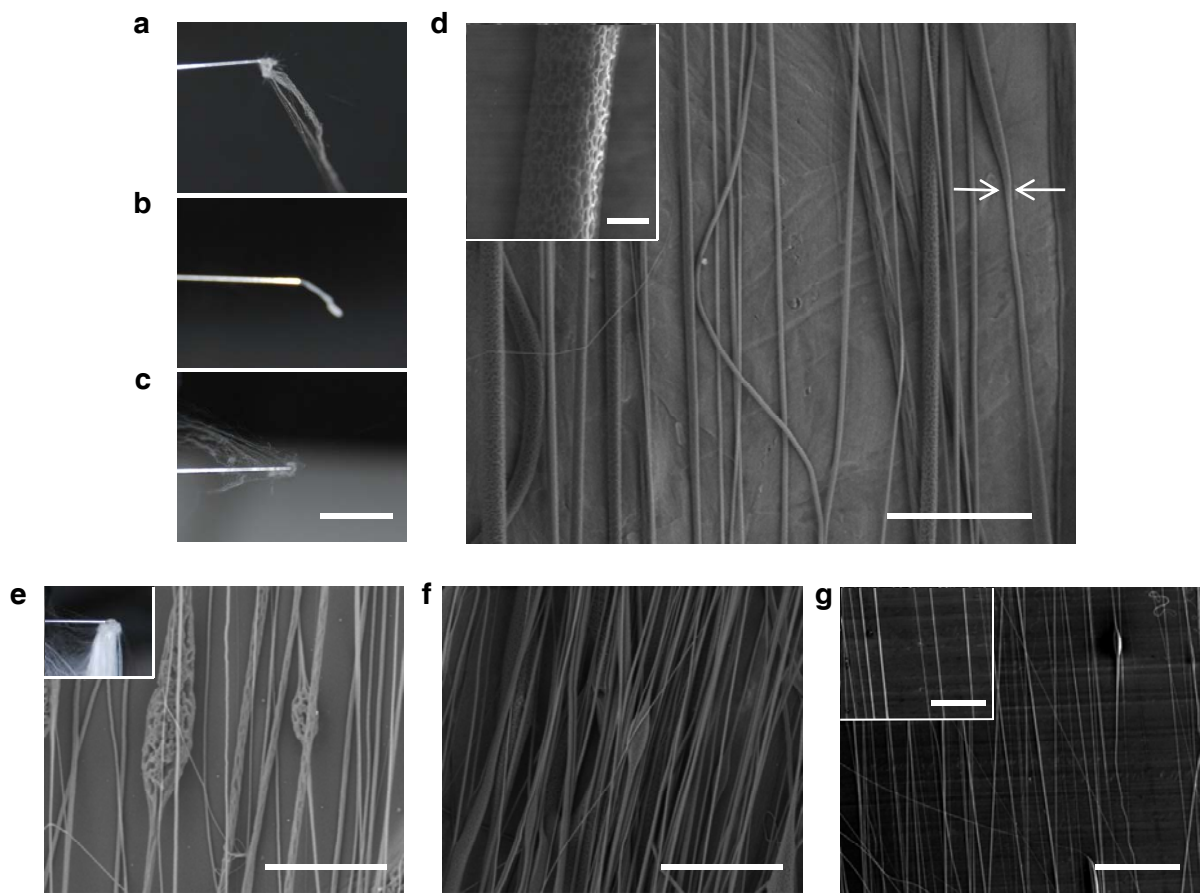

**Supplementary Figure S7 │ Electrospinning experiments with low boiling solvents.** Photographs of needle clogging during electrospinning of PVDF-TrFe dissolved in Acetone (**a**), THF (**b**) and MEK (**c**) at a polymer/solvent concentration of 21% (w/w). Scale bar = 1 cm. (**d**) SEM micrograph of PVDF-TrFe fibers electrospun from a MEK solution (21%, w/w) prior to needle clogging. Arrows indicate an example of strong diameter variation along the fiber length. Scale bar 10 μm. Inset: magnified view of a single fiber to reveal the level of surface porosity . Scale bar 1 μm. (**e**) SEM micrograph of PVDF-TrFe fibers made by using ACE (12%, w/w). Scale bar 10 μm. Inset: photograph of needle clogged after 15 minutes of electrospinning. (**f**) SEM micrograph of PVDF-TrFe fibers made by using THF (12%, w/w). Scale bar 10 μm. (**g**) SEM micrograph of PVDF-

TrFefibers made by using MEK (17%, w/w). Scale bar 100 μm. Inset: Magnified SEM micrograph of the same sample. Scale bar: 40 μm.

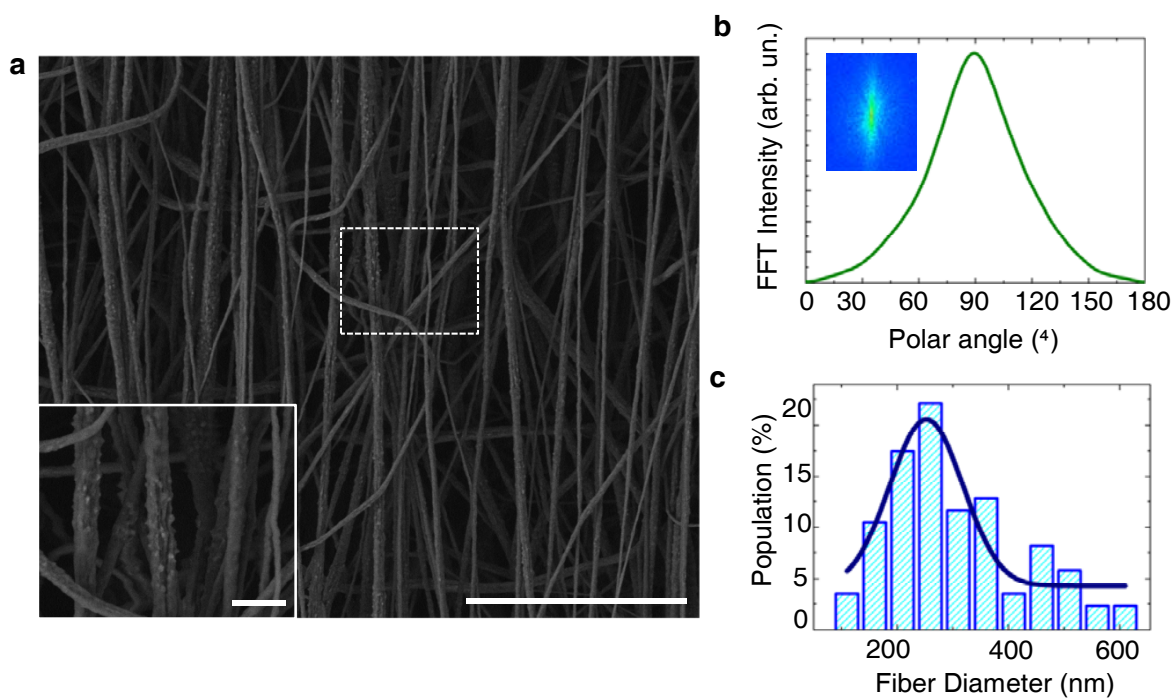

**SupplementaryFigure S8│Aligned array of PVDF nanofibers.** (**a**) SEM micrographs of PVDF fiberselectrospun using the same experimental conditions as those for PVDF-TrFe. Scale bar: 10 μm. Inset: magnified view of the region indicated by a dashed square. Scale bar: 1 μm. (**b**) Radial intensity distribution vs detection angle (0-180°) for aligned arrays of PVDF fibers. The full width at half maximum is 51°. Inset: 2D FFT image.(**c**) Typical PVDF fiber diameter distribution and fit to a Gaussian shape (solid line).

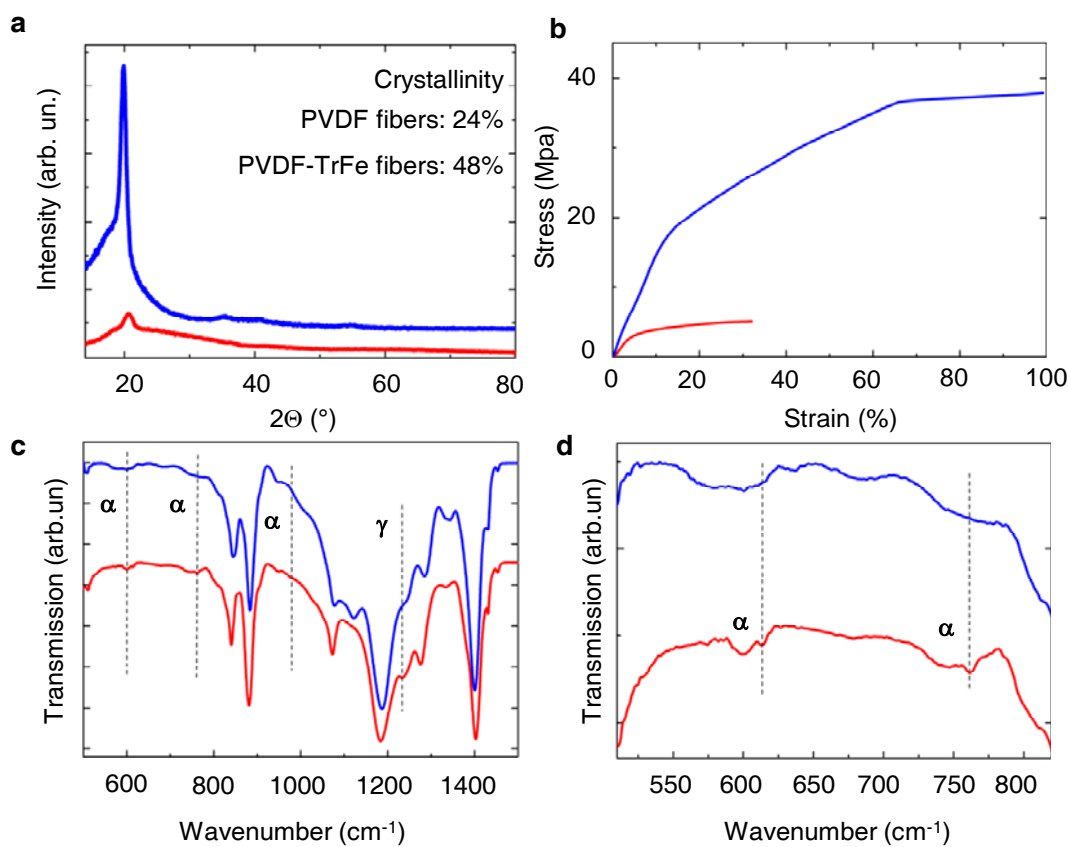

**Supplementary Figure S9│Characterization of aligned arrays of PVDF nanofibers.** (**a**) X-ray diffraction patterns (XRD; Cu radiation with wavelength = 0.15418 nm) from aligned fibers of PVDF (red line) and PVDF-TrFE (blue line). The data are vertically shifted to facilitate comparison. (**b**) Typical stress-strain curves measured from aligned arrays of fibers of PVDF (red line) and PVDF-TrFe (blue line). (**c**) FTIR spectra of PVDF (red line) and PVDF-TrFe (blue line) aligned fibers. (**d**) Magnified view of the high energy region of the FTIR spectra, highlighting non-polar α bands at 612 and 762 cm$^{-1}$.

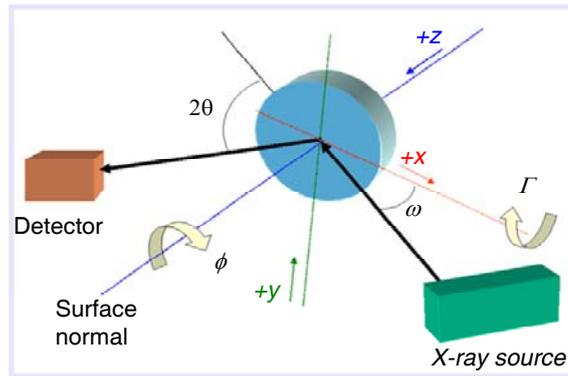

**Supplementary Figure S10│Diagram of X-ray diffraction instrument.** X-ray analysis was carried out by performing 2θ/ω scans, where ω is the angle of incidence of X-ray beam relative to the sample surface and 2θ is the diffraction angle. The diffraction plane is defined by the incident and diffracted directions of the X-ray beams. Measurements performed at azimuthal rotation $\phi = 0$ where the fibers length parallel to the diffraction plane. $\Gamma$ indicatethe tilt angle.

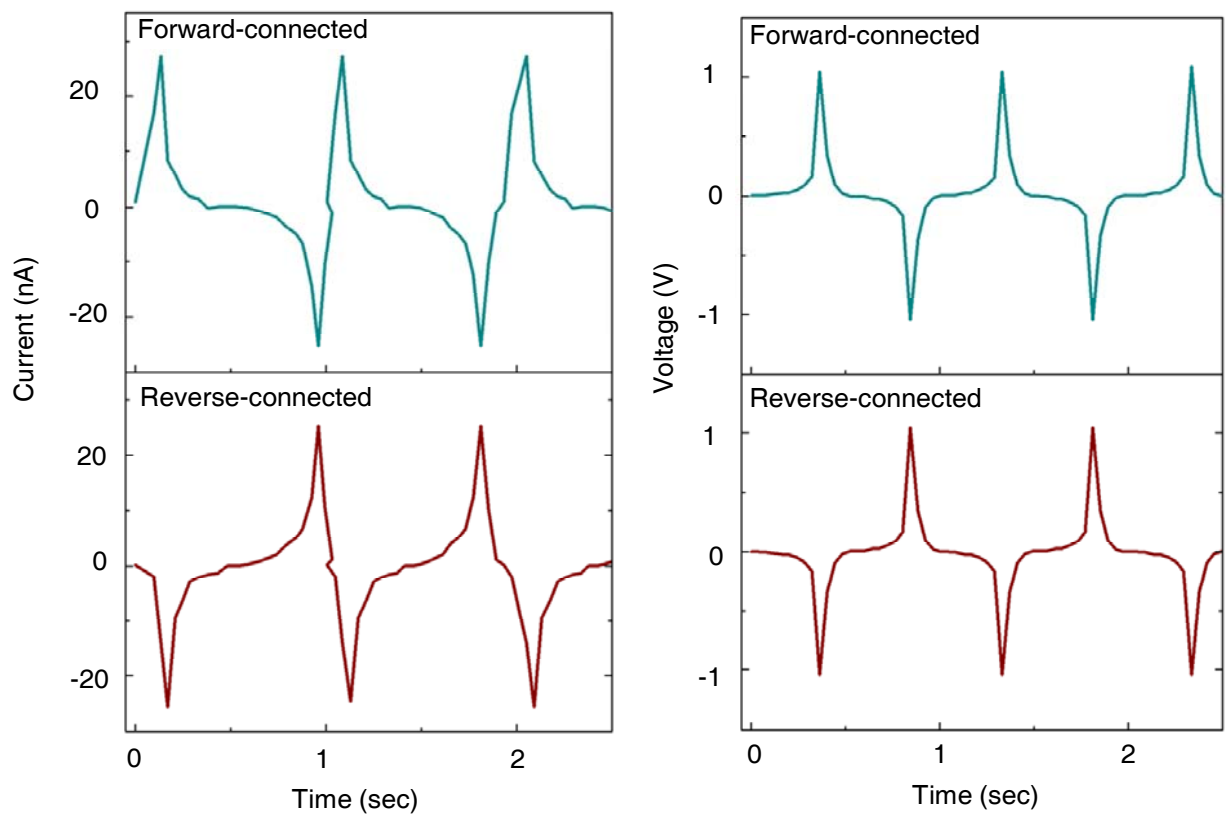

**Supplementary Figure S11│Current and voltage measurements with reversed connections.** Short-circuit current and open circuit voltage measurements output of a typical nanofiber array device with forward (cyane blue line) and reversed (red wine line) connection to the measurement system. Connections showed reversal in signal polarity as expected.

**Supplementary Note 1. X-ray Photoelectron Spectra (XPS)**

X-ray photoelectron spectroscopy (XPS) is a quantitative spectroscopic technique that was used to calculate copolymerization ratio of PVDF-TrFEfiber array. XPS spectra were collected using a Kratos Axis ULTRA X-ray photoelectron spectrometer with monochromatic Al Kα excitation, 120 W (12kV, 10 mA). In order to reduce the effects of surface charging, the monochromatic source was operated using a bias voltage of 100 V. Data were collected using the low magnification (FOV1) lens setting with a 2 mm aperture (200 μm analysis area) and charge neutralizer settings of 2.1 A filament current, 2.1 V charge balance and 2 V filament bias. Survey spectra were collected at a pass energy of 160 eV and high resolution spectra were collected using a pass energy of 40 eV. The data were fitted with Gaussian-Lorentzian line shapes. The binding energy scale was referenced to the aliphatic C 1s line at 285.0 eV.

The copolymerization ratio was calculated by two different methods:

1) using Atomic concentrations calculated by curve fitting

Considering, the total % of Fluorine, % F= 16.03+36.73= 52.76 and that of Carbon,

% C= 2.39+17.22+6.43+20.44+0.77=47.25

The atomic ratio is $\frac{C}{F} = 0.90$             (i)

By taking into account the molecular formula of PVDF, $[(CH_2 - CF_2)_x]$ and TrFe, $[(CHF - CF_2)_{(1-x)}]$, the atomic ratio can be represented as,

$$\frac{C}{F} = \frac{2x + 2(1-x)}{2x + 3(1-x)} = \frac{2}{3-x} \quad (ii)$$

Equaling (i) and (ii),

$$\frac{C}{F} = 0.90 = \frac{2}{3-x}$$

Therefore, the fraction of PVDF, x= 0.78 and the fraction of TrFe, 1-x = 0.22

2) using peak area ratios calculated by using C1s curve fitting:

| CharacteristicPeak | DOS (BE) | Area % | Normalized to 100% |
|---|---|---|---|
| PVDF | 286.7 | 36.45 | 72.81 |
| TrFE | 289.0 | 13.61 | 27.17 |

The fraction of PVDF equals to 0.73, while the fraction of TrFe is 0.27

From the XPS spectra, the stoichiometric ratio (in atomic percent) of two elements, C and F, in the P(VDF–TrFe) film is determined. Co-polymerization ratio of PVDF:TrFe (0.73:0.27) from XPS results shows excellent agreement with actual co-polymer ratio of commercial PVDF:TrFE (0.75:0.25).

**Supplementary Note 2. Piezoelectric analysis of PVDF-TrFefiber arrays under compression**

The constitutive model of piezoelectric materials gives the relations among the stress $\sigma_{ij}$, strain $\varepsilon_{ij}$, electric field $E_i$ and electric displacement $D_i$ as

$$\begin{Bmatrix} \sigma_{11} \\ \sigma_{22} \\ \sigma_{33} \\ \sigma_{23} \\ \sigma_{31} \\ \sigma_{12} \end{Bmatrix} = \begin{bmatrix} c_{11} & c_{12} & c_{13} & 0 & 0 & 0 \\ c_{12} & c_{11} & c_{13} & 0 & 0 & 0 \\ c_{13} & c_{13} & c_{33} & 0 & 0 & 0 \\ 0 & 0 & 0 & c_{44} & 0 & 0 \\ 0 & 0 & 0 & 0 & c_{44} & 0 \\ 0 & 0 & 0 & 0 & 0 & (c_{11}-c_{12})/2 \end{bmatrix} \begin{Bmatrix} \varepsilon_{11} \\ \varepsilon_{22} \\ \varepsilon_{33} \\ 2\varepsilon_{23} \\ 2\varepsilon_{31} \\ 2\varepsilon_{12} \end{Bmatrix} - \begin{bmatrix} 0 & 0 & e_{31} \\ 0 & 0 & e_{31} \\ 0 & 0 & e_{33} \\ 0 & e_{15} & 0 \\ e_{15} & 0 & 0 \\ 0 & 0 & 0 \end{bmatrix} \begin{Bmatrix} E_1 \\ E_2 \\ E_3 \end{Bmatrix}, \quad (S1)$$

$$\begin{Bmatrix} D_1 \\ D_2 \\ D_3 \end{Bmatrix} = \begin{bmatrix} 0 & 0 & 0 & 0 & e_{15} & 0 \\ 0 & 0 & 0 & e_{15} & 0 & 0 \\ e_{31} & e_{31} & e_{33} & 0 & 0 & 0 \end{bmatrix} \begin{Bmatrix} \varepsilon_{11} \\ \varepsilon_{22} \\ \varepsilon_{33} \\ 2\varepsilon_{23} \\ 2\varepsilon_{31} \\ 2\varepsilon_{12} \end{Bmatrix} + \begin{bmatrix} k_{11} & 0 & 0 \\ 0 & k_{22} & 0 \\ 0 & 0 & k_{33} \end{bmatrix} \begin{Bmatrix} E_1 \\ E_2 \\ E_3 \end{Bmatrix}. \quad (S2)$$

Consider a uniaxial compressionapplied along the $x_1$ direction (normal to fibers, Fig. 3a) instantaneously at time $t=0$, i.e., i.e., $\sigma_{11}=-pH(t)$, where $H$ is the Heavyside step function $H(t)=\begin{cases} 0 & \text{for } t<0 \\ 1 & \text{for } t\geq 0 \end{cases}$. For $\varepsilon_{22}=\varepsilon_{33}\approx 0$ because the P(VDF-TrFE) fiber arrays (elastic modulus ~ 200 MPa) are bonded to the much thicker and stiffer plastic substrate (elastic modulus 2.5 GPa), Eqs. (S1) and (S2) give $-pH(t)=c_{11}\varepsilon_{11}-e_{31}E_3$ and $D_3=e_{31}\varepsilon_{11}+k_{33}E_3$, respectively, where $D_3$ is the electric displacement along the poling direction, and the electric field $E_3$ is related to the voltage $V$ and effective contact length $L_{eff}$ by $E_3=V/L_{eff}$. Elimination of $\varepsilon_{11}$ from these two equations yields the electric displacement $D_3=-\bar{d}pH(t)+(\bar{k}V/L_{eff})$, where $\bar{d}=e_{31}/c_{11}$ and $\bar{k}=k_{33}+e_{31}^2/c_{11}$. The voltage $V$ and current $I=-h_{PVDF-TrFe}w_{PVDF-TrFe}\dot{D}_3$ are also related by the resistance $R$ of the voltmeter, $V=IR$. These yield the coupling equation of the voltage

$$\frac{dV}{dt}+\frac{L_{eff}}{\bar{k}h_{PVDF-TrFe}w_{PVDF-TrFe}R}V=\frac{\bar{d}L_{eff}}{\bar{k}}\delta(t), \quad (S3)$$

where $\delta(t)$ is the delta function. For the initial condition $V(t=0^-)=0$, the solution of Eq. (S3) is $V=(\bar{d}L_{eff}/\bar{k})p\exp\left[-L_{eff}t/(\bar{k}Rh_{PVDF-TrFe}w_{PVDF-TrFe})\right]$. The maximum value is given in Eq.(1) in the main text.

**Supplementary Note 3. Piezoelectric analysis of PVDF-TrFefiber arrays under bending**

### 3.1 Mechanics analysis

For the out-of-plane displacement $w = A[1+\cos(2\pi x_3/L_{PI})]/2$ in plane-strain analysis ($\varepsilon_{22} = 0$), the bending energy in flexible polymide support (PI) is $(EI/2)\int(w'')^2 ds$, where $w''$ and $EI$ are the curvature and plane-strain bending stiffness of PI substrate, respectively, and the integration is over the length of PI substrate. The membrane energy can be obtained following the same approach of Song et al.[39] Minimization of total energy (sum of bending and membrane energies) gives the amplitude $A$ as

$$A = \frac{2}{\pi}\sqrt{L_{PI}\cdot\Delta L - \frac{\pi^2 t_{PI}^2}{3}} \approx \frac{2}{\pi}\sqrt{L_{PI}\cdot\Delta L}, \tag{S4}$$

where the last approximation holds when the compression of PI substrate $\Delta L \approx t_{PI}^2/L_{PI}$ as in the experiment. The strain in the fiber arrays is obtained from the curvature and the distance between the mid-planes of PI substrate and fiber arrays as

$$\varepsilon_{33} = \frac{\pi^2(h_{PI} + h_{PVDF-TrFe})A}{L_{PI}^2}\cos\left(\frac{2\pi x_3}{L_{PI}}\right). \tag{S5}$$

The pure bending of PIsubstrate, together with the traction-free condition on the surface of the fiber arrays, gives $\sigma_{11} = 0$ in the fiber arrays. Its substitution into Eq. (S1) yields $\varepsilon_{11} = -(c_{13}/c_{11})\varepsilon_{33} + (e_{31}/c_{11})E_3$. The electric field $E_3$ is then obtained from Eq. (S2) as

$$E_3 = -\frac{\bar{e}}{\bar{k}}\varepsilon_{33} + \frac{1}{\bar{k}}D_3, \tag{S6}$$

where the electric displacement $D_3$ is spatially invariant (established from the charge equation $dD_3/dx_3 = 0$) and is to be determined.

### 3.2 Current

The voltage across the length of the fiber arrays $L_{PVD-TreF}$ is zero, $\int_{-L_{PVDF-TrFe}/2}^{L_{PVDF-TrFe}/2} E_3 dx_3 = 0$, after the fiber arrays is connected to an ampere meter. Together with Eqs. (S5) and (S6), it gives the electric displacement

$$D_3 = \frac{2\bar{e}(h_{PI} + h_{PVDF-TrFe})}{L_{PVDF-TrFe}} \sqrt{\frac{\Delta L}{L_{PI}}} \sin\left(\pi \frac{L_{PVDF-TrFe}}{L_{PI}}\right). \quad (S7)$$

The electric charge is the product of $-D_3$ and the cross section area $h_{PVDF\text{-}TrFe} w_{PVDF\text{-}TrFe}$ of the fiber arrays. Its rate of change gives the current

$$I = -\frac{\bar{e}(h_{PI} + h_{PVDF-TrFe})h_{PVDF-TrFe}w_{PVDF-TrFe}}{L_{PVDF-TrFe}\sqrt{L_{PI}\Delta L}} \frac{d\Delta L}{dt} \sin\left(\pi \frac{L_{PVDF-TrFe}}{L_{PI}}\right). \quad (S8)$$

For the representative $\Delta L$ in the main text, the maximum current in Eq. (2) is obtained.

### 3.3 Voltage

Let $V$ denote the voltage across the length of the fiberarrays $L_{PVDF\text{-}TrFe}$ after the fiberarrays is connected to a voltmeter. This requires $\int_{-L_{PVDF-TrFe}/2}^{L_{PVDF-TrFe}/2} E_3 dx_3 = V$, which, together with Eqs. (S5) and (S6), gives the electric displacement

$$D_3 = \frac{2\bar{e}(h_{PI} + h_{PVDF-TrFe})}{L_{PVDF-TrFe}} \sqrt{\frac{\Delta L}{L_{PI}}} \sin\left(\pi \frac{L_{PVDF-TrFe}}{L_{PI}}\right) + \frac{\bar{k}}{L_{PVDF-TrFe}} V. \quad (S9)$$

Similar to Supplementary Note 3.2, the rate of change of Eq. (S9) gives the current

$$I = -\frac{\bar{e}(h_{PI} + h_{PVDF})h_{PVDF-TrFe}w_{PVDF-TrFe}}{L_{PVDF-TrFe}\sqrt{L_{PI}\Delta L}} \frac{d\Delta L}{dt} \sin\left(\pi \frac{L_{PVDF-TrFe}}{L_{PI}}\right) - \frac{\bar{k} h_{PVDF-TrFe} w_{PVDF-TrFe}}{L_{PVDF-TrFe}} \frac{dV}{dt}. \quad (S10)$$

Its relation with the resistance $R$ of the voltmeter leads to Eq. (3) and the maximum voltage in Eq. (4).

**Supplementary Note 4. Pyroelectric measurements**

Pyroelectric effects were investigated by measuring both the current and voltage output of a nanofiber array device using the continuous heating/cooling method at zero dc voltage and at a constant heating/cooling rate between 1 to 6 K/min. A small stage (All Electronics Corp.; 40x44 mm, which closely matches the lengths of the fibers) was used to investigate effects of temperature on just the fiber region of the device as shown in Figure S4a. We also used a slightly larger stage (Monster; 62x62 mm) as shown in Figure S4b to examine the effects of temperature on the entire device, including the electrode contacts. The heating rates and temperature ranges correspond, roughly, to those in references 40, 41, 42. Figure S4 provides a schematic illustration of the setups. Throughout the experiments, the sample temperature was measured with a NiAl thermocouple and a Model HH21 Omega microprocessor thermometer. The pyroelectric coefficientcan be measured by the Lang-Steckel method[41].The governing equation for the pyroelectric effect is

$$D_3 = \alpha T + k_{33} E_3, \quad (S11)$$

where $\alpha$ is the pyroelectric coefficient, and $T$ is the change of the temperature. For current measurements using an ammeter, the voltage across the fiber length, and therefore the electric field $E_3$, are zero such that $D_3 = \alpha T$. The rate of the change in temperature creates a currentgiven by $I(t) = -\alpha h_{PVDF-TrFe} w_{PVDF-TrFe} dT/dt$, which can be used to measure the pyroelectric coefficient as

$$\alpha = -\frac{I(t)}{h_{PVDF-TrFe} w_{PVDF-TrFe} dT/dt}. \quad (S12)$$

For the measurement results in Figure S5 $I(t)/(dT/dt) = 10.9 \ pC/K$, and $h_{PVDF\text{-}TrFe}$=20 μm and $w_{PVDF\text{-}TrFe}$= 8 mm, Eq. (S12) gives the pyroelectric coefficient as $\alpha$=-68μC/(m$^2$K), which is in the same range as values reported for thin films of PVDF-TrFe films (-20μC/m$^2$K) [42]. At a constant heating rate of 2.5 K/min, the pyroelectric signals follow the derivative of the temperature, with maximum voltages of 1.9 mV and 1.3 mV when contacts are on top of and outside of the heater, respectively. The responses are symmetrical with respect to heating and cooling, as expected. The magnitudes of these voltages are nearly 50 times smaller than the smallest piezovoltages reported in the manuscript (i.e. those that result from tests at 0.1 Pa).The corresponding pyroelectric analysis is reported in Supplementary Note 5.

**Supplementary Note 5. Pyroelectric analysis**

For voltage measurement of the fiber arrays via a voltmeter, Eq. (S11) still holds. The electric field $E_3$ and electric displacement $D_3$ are related to the voltage $V$ and current $I$ by $E_3 = V/L_{PVDF-TrFe}$ and $I = -h_{PVDF-TrFe} w_{PVDF-TrFe} \dot{D}_3$, respectively, where $L_{PVDF-TrFe}$ is the length of fiber arrays and $h_{PVDF-TrFe} w_{PVDF-TrFe}$ is the cross section area. The voltage $V$ and current $I$ are also related by the resistance $R$ of the voltmeter, $V=IR$. These give the equation for V as

$$\frac{dV}{dt} + \frac{L_{PVDF-TrFe}}{k_{33} R h_{PVDF-TrFe} w_{PVDF-TrFe}} V = -\frac{\alpha L_{PVDF-TrFe}}{k_{33}} \frac{dT}{dt}. \quad (S16)$$

For the initial condition $V(t=0)=0$, the above equation has the solution

$$V = -\frac{\alpha L_{PVDF-TrFe}}{k_{33}} \int_0^t \frac{dT}{d\tau} e^{\frac{L_{PVDF-TrFe}}{k_{33} R h_{PVDF-TrFe} w_{PVDF-TrFe}}(\tau - t)} d\tau. \quad (S17)$$

For $L_{PVDF-TrFe}$=40 mm, thickness $h_{PVDF}$=20 μm, width $w_{PVDF-TrFe}$=8 mm, and the measured pyroelectric coefficient $\alpha = -68$ μC$/(m^2 K)$ and resistance of the voltmeter $R$=4MΩ in experiments, the maximum voltage is 1.81 μV for the measured temperature $T(t)$ in Fig. S6 and the dielectric constant $k_{33}$=5.31*10$^{-11}$ F/m [38]. In fact, for the normalized time $L_{PVDF-TrFe} t/(S k_{33} R) \Box 1$, Eq.(S17) can be simplified to Eq. (5) in the main text, which also gives the maximum voltage 1.81 V, and agrees well with maximum voltage of 1.9 V when the PVDF-TrFe fiber arrays are in contact with the top of the heater.

It should be pointed out that different voltmeters were used for pyroelectric and piezoelectric measurements, with resistance of the voltmeter R=4 MΩ and 70 MΩ, respectively due to different sensing range of voltmeters. Even for R=70 MΩ and heating rate 6.5 K/min, the pyroelectric voltage is still smaller than the piezoelectric voltage even at the lower end of the range of pressure sensitivity (0.1 Pa, in the type of tests reported here).

**Supplementary Note 6. Comparison between PVDF-TrFe and PVDF fibers**

For purposes of comparison, PVDF fibers were formed with the same experimental conditions used for PVDF-TrFe. In particular, PVDF (Sigma Aldrich) was dissolved in a 3:2 volume ratio of dimethylformamide/acetone at a polymer/solvent concentration of 21% w/w. A potential of 30 kV was applied between a nozzle tip with inner diameter of 200 µm, fed by a syringe pump at a flow-rate of 1 mL/hr, and a collector at a distance of 6 cm. The collector disk rotated at angular speed of 4000 rpm, corresponding to linear speeds > 16 m/s at the collector surface. The fibers were collected on aluminium strips with widths of 0.8 cm and lengths of 25 cm. The morphological, crystallographic and mechanical properties of PVDF fibers were measured and compared with those of PVDF-TrFe. The most immediate, striking difference from PVDF-TrFe was that the PVDF fibers were not sufficiently robust to exist as stable, free-standing films. In nearly all cases the process of detaching the PVDF fibers from the aluminium foil mechanically destroyed the samples by fracture and tearing. Furthermore, we found that PVDF fibers are poorly aligned (Figure S8a). 2D FFT analysis indicates that the full width at half maximum of the radial intensity distribution of the elliptical profile is 51°, which is more than three times larger than that of the PVDF-TrFe arrays (Figure S8b). Although the average fiber diameters (250 nm) and the corresponding distribution ofdiameters are comparable (Figure S8c), the surfaces of the PVDF fibers are rough, with protrusions that have characteristic dimensions of ~100 nm (Inset of Figure S9a). X-ray diffraction (XRD) patterns indicate that PVDF fibers exhibit 24% crystallinity, which is two times lower than that of PVDF-TrFefibers (Figure S9a) and the non-polar α (612,762, 976 cm$^{-1}$) and γ (1234 cm$^{-1}$) phases are here clearly distinguishable by FTIR (Figure S9c and d). The mechanical properties were investigated using dynamicmechanicalanalysis (DMA Q800, TA Instruments, New Castle, DE), in tensile mode at constant temperature (25°C). At least three different fibrous specimens for each polymer were tested. Specimen dimensions were approximately 8.0×12.0 mm (width×length) with thicknesses between 20 and 40 µm. The stress-straincurveswererecordedwith a ramp/rate of 0.5 N/min (upto 18 N). Typical stress-strain curves of PVDF and PVDF-TrFefibers are reported in Figure S9b. The measured Young Modulus is 168 ±8 MPa for PVDF-TrFe and 86 ±11 MPa for PDVF fibers respectively. The maximum elongation of PVDF-TrFEfibers is ~100% (corresponding to a tensile strength of about 40 MPa) while that of PVDF fibers is ~32% (corresponding to a tensile strength of about 5 MPa). These results clearly indicate that PVDF-TrFefibers exhibit superior mechanical properties compared to PVDF fibers. Such differences are critically important to use in the classes of devices described in our manuscript. In addition, the formation of inter-fibers joints or adhesion points further increase both the tensile strength and the elongation path[44].

**Supplementary methods . PVDF-TrFe fibers made with low-boiling solvents**

PVDF-TrFe was dissolved in three different low-boiling point ($T_b$) solvents: Acetone (ACE, $T_b$: 57°C), Tetrahydrofuran (THF, $T_b$: 66°C) and methylethylketone (MEK, $T_b$: 80°C) at different polymer/solvent concentrations in the range 12-21% (w/w). At the highest concentration, needle clogging occurs almost instantaneously with ACE and THF (Figure S7 a and b respectively), and after tens of seconds with MEK (Figure S7c). In this last case, collected fibers take the form ofisolated strands (about $7 \times 10^2$ roughly parallel fibers per mm) with an average diameter ~570 nm. Such fibers have non-uniform morphologies, with both flat and porous surfaces, and in short, discontinuous segments due to the presence of necks and beads (Figure S7d).

At lower concentrations, the maximum electrospinning time could be extended to ~15 minutes with ACE (corresponding to 300 μL of solution at the lowest concentration of 12% w/w) before needle clogging stopped irreversibly the electrospinning process. Fibers in this case appear in the form of isolated strands ($2 \times 10^3$ fibers per mm) with an average diameter of ~340 nm. As with MEK, fiber continuity is interrupted due to the formation of numerous beads (width of 2-8 μm and length of 5-20 μm) and the surface morphology of the fibers is inhomogeneous. In general, such discontinuities are observed at any solution concentration (Figure S7e).

Using THF, electrospinning was not interrupted by needle clogging, but the produced fibers are, nevertheless, discontinuous with beads and a large variety of surface morphologies. Fibers exhibit an average diameter of ~570 nm, and a density below $1 \times 10^3$ fibers per mm for solutions at 12% polymer/solvent (Figure S7f). Finally in case of MEK beads and surface discontinuity can be strongly reduced but the resulting fibers are still in the form of isolated strands ($1 \times 10^2$ fibers per mm) with an average diameter of 590 nm even after 1 hour of continuous spinning (Figure S7g).

**Supplementary References**